 \documentclass[aps,prb,showpacs,superscriptaddress,byrevtex,showpacs,amsmath,amssymb,twocolumn]{revtex4}
\usepackage{epsfig}
\usepackage{dcolumn}
\usepackage{bm}
\usepackage{amsmath}
\usepackage{amsfonts}
\usepackage{latexsym}
\usepackage{amssymb}
\usepackage{color}
\usepackage{hyperref}
\usepackage[normalem]{ulem}

\usepackage{fancyhdr}
\usepackage{amsbsy}
\usepackage{amsmath}
\usepackage{epsfig}
\usepackage{amssymb}
\usepackage{pifont}
\usepackage{textcomp}
\usepackage{gensymb}
\usepackage{pifont}
\usepackage{enumerate}
\usepackage{color}
\usepackage{latexsym}
 	\definecolor{ballblue}{rgb}{0.13, 0.67, 0.8}
\definecolor{bazaar}{rgb}{0.6, 0.47, 0.48}
\definecolor{cadmiumgreen}{rgb}{0.0, 0.42, 0.24}

\begin{document}	

\title{Spin and charge transport in ferromagnet-superconductor-ferromagnet heterostructures: Stoner versus spin mass mismatch mechanism}

\author{Paola Gentile}
\affiliation{Consiglio Nazionale delle Ricerche CNR-SPIN, I-84084 Fisciano (Salerno), Italy}
\affiliation{Dipartimento di Fisica "E.R. Caianiello", Universit\`a degli Studi di Salerno, I-84084 Fisciano
(SA), Italy}

\author{Marilena Catapano}
\affiliation{Universit\`a degli Studi di Salerno, I-84084 Fisciano (SA), Italy}

\author{Nicola De Vivo}
\affiliation{Intesa Sanpaolo - Direzione Rischi Finanziari e di Mercato, Piazza Paolo Ferrari 10,
20121 Milano}

\author{Mario Cuoco}
\affiliation{Consiglio Nazionale delle Ricerche CNR-SPIN, I-84084 Fisciano (Salerno), Italy}
\affiliation{Dipartimento di Fisica "E.R. Caianiello", Universit\`a degli Studi di Salerno, I-84084 Fisciano
(SA), Italy}

\author{Alfonso Romano}
\affiliation{Dipartimento di Fisica "E.R. Caianiello", Universit\`a degli Studi di Salerno, I-84084 Fisciano
(SA), Italy}
\affiliation{Consiglio Nazionale delle Ricerche CNR-SPIN, I-84084 Fisciano (Salerno), Italy}

\author{Canio Noce}
\affiliation{Dipartimento di Fisica "E.R. Caianiello", Universit\`a degli Studi di Salerno, I-84084 Fisciano
(SA), Italy}
\affiliation{Consiglio Nazionale delle Ricerche CNR-SPIN, I-84084 Fisciano (Salerno), Italy}

\date{\today}

\begin{abstract}
We study transport phenomena through a ballistic ferromagnet-superconductor-ferromagnet (F/S/F) junction, comparing the case in which the ferromagnetic order in the two F layers is of the standard Stoner type with the case where it is driven by a spin mass mismatch (SMM). It is shown that the two mechanisms lead to a different behavior in the charge and the spin conductances, especially when compared to the corresponding non-superconducting ferromagnet-normal-ferromagnet (F/N/F) junctions. In particular, when the injected current is perpendicular to the barrier, for high barrier transparency and large magnetization of the F layers, the large mass mismatch gives rise to an enhancement of both low-bias charge and spin conductances of the F/S/F junction, which is not observed in the equal-mass case. 
When all the allowed injection directions are considered, the low bias enhancement of the charge conductance for SMM leads still holds for high barrier transparency and large magnetization of the F layers. However, in the case of non-transparent interfaces, spin transport with SMM ferromagnets exhibits an opposite sign response with respect to the Stoner case at high biases for all magnetization values, also manifesting a significant amplification induced by superconductivity at the gap edge. The above mentioned differences can be exploited to probe the nature of the electronic mechanism underlying the establishment of the ferromagnetic order in a given material.   
\end{abstract}

\vspace{2pc}


\maketitle

\section{Introduction}
Heterostructures made of ferromagnetic (F) and superconducting (S) alternating layers exhibit a variety of peculiar phenomena occurring at the nanoscale range of layer thicknesses~\cite{Buzdin05,Bergeret05,Linder15}. Thanks to the great progress in the preparation of high-quality hybrid F/S systems achieved in the last years, their properties have been deeply investigated in view of the design of new devices susceptible of relevant applications in the field of electronics and spintronics~\cite{Linder15,Hirohata14}. The interest in the above-mentioned systems is however not limited to this context. Under specific conditions, their behavior may provide relevant information on the type of ferromagnetism characterizing the F layer as well as on the symmetry of the order parameter in the superconductor, allowing to distinguish among the various possible unconventional pairing states~\cite{Kirtley}.

\begin{figure}[htbp]
\includegraphics[width=0.9\columnwidth, angle=0]{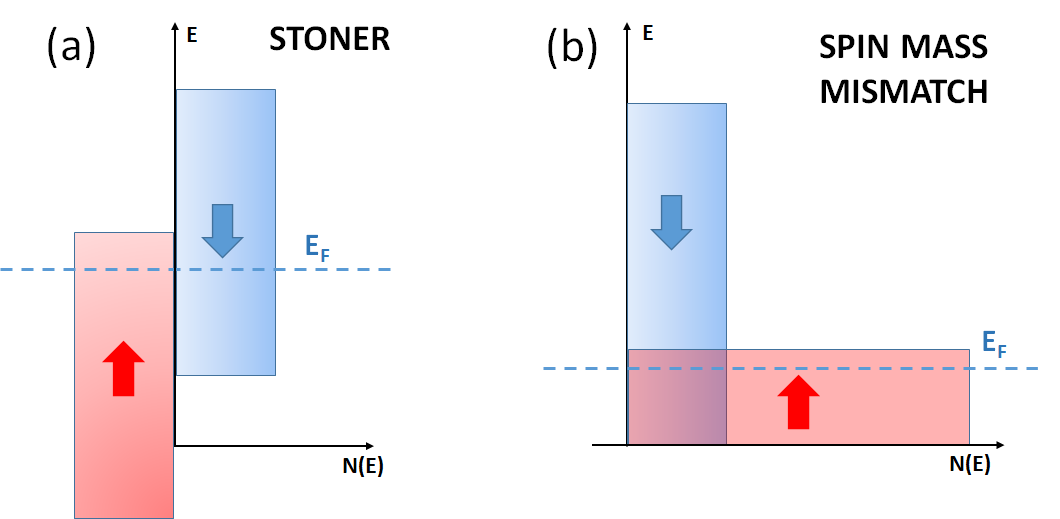}
\caption{Density of states for spin-up and spin-down electrons in the Stoner (a) and in the spin mass mismatch (SMM) case (b). }
\label{DOS} 
\end{figure}

Most of the relevant effects arising in F/S structures, such as the spatial oscillations of the electronic density of states or the nonmonotonic dependence of the critical temperature on the ferromagnet layer thickness, are ultimately related to the damped oscillatory behavior characterizing the propagation of the Cooper pair wave function from the superconductor to the ferromagnet~\cite{Demler97}. This is in turn due to the formation of Cooper pairs with a finite center-of-mass momentum originating from the presence of the exchange field~\cite{FFLO}.
As far as transport is concerned, it is well known that in a N/S junction, where N denotes a normal metal, for energies below the superconducting gap $\Delta$, conduction is only possible via Andreev reflection (AR) processes by which two electrons with opposite spins, one above and the other one below the Fermi energy, incident from the non-superconducting layer, are transferred in the superconductor as a Cooper pair~\cite{Andreev64}. This leaves holes in the normal system which give rise to a parallel conduction channel, in this way leading to a doubling of the normal-state conductance for $eV<\Delta$ ($V$ is the applied voltage)~\cite{Soulen98}. When the normal metal is replaced by a conventional Stoner ferromagnet, the relative shift of the density of states for spin-up and spin-down electrons caused by the exchange interaction (Fig.~\ref{DOS}) can be large enough that an electron with, say, spin-up incident on the interface finds no spin-down partner to form a Cooper pair able to move to the superconducting layer~\cite{Beenakker95}. Andreev reflections at the interface are then blocked so that only single-particle excitations contribute to the conductance. As a result, the higher is the exchange interaction in the ferromagnet, the stronger is the conductance suppression in the subgap energy range~\cite{Vasko98}.

In F/S junctions, Andreev reflections occur as local processes at the superconductor interface and produce a Cooper pair in the superconductor. In multiterminal F/S hybrid structures where the thickness of the S layer is of the order of the BCS superconducting coherence length of the material, they can also manifest themselves as a nonlocal process~\cite{Byers95}, referred to as crossed Andreev reflection (CAR). In such a case, the retroreflection of the hole from an AR process, resulting from an incident electron at energies less than the superconducting gap at one lead, occurs in the second ferromagnetic lead with the same charge transfer as in a normal AR process of a Cooper pair in the superconductor. For CAR to occur, electrons of opposite spin must exist at each non superconducting electrode (so as to form the pair in the superconductor). Therefore such processes are expected to be strongly suppressed in F/S/F junctions with parallel alignment of the F polarizations, while they can survive even at strong polarization in the case of F layers having opposite magnetization alignment. 
The reverse process of the CAR produces spatially separated entangled states of electrons by splitting Cooper pairs from the superconducting condensate into the two external leads~\cite{Recher2001,Herrmann2010}. In trilayer structures, CARs usually compete with other transport processes, such as the normal reflections, the local Andreev reflections, and the elastic cotunneling, i.e. the quantum mechanical tunneling of electrons between the external leads via an intermediate state in the superconductor.

Generally speaking, F/S/F trilayer structures offer a rich playground to investigate the interplay between superconductivity and ferromagnetism. For instance, in such structures theory predicts~\cite{Tagirov99} that for parallel alignment of the magnetizations in the two ferromagnetic layers, the superconducting critical temperature is lower than in the case of antiparallel alignment, and can even be zero. The system can thus behave as a spin valve where superconductivity can be switched on and off by reversing the field direction in one of the two magnetic layers. However, it has been experimentally verified~\cite{Gu02,Potenza05,Moraru06,Nowak08} that as soon as collinear, i.e. parallel or antiparallel, configurations are considered, the critical temperature shift is of the order of millikelvins, i.e. relatively small compared to the theory predictions. The smallness of this spin valve effect has been supposed to be due to a non-optimal choice of the layer thickness and/or the selected layer material~\cite{Kehrle12}. Actually, in the collinear case most of the experiments have been performed on systems not satisfying the condition $\xi_\textsc{S}/d_\textsc{S}\ge 1$, $\xi_\textsc{S}$ and $d_\textsc{S}$ being the coherence length and the thickness of the superconducting layer, respectively, which has been theoretically demonstrated~\cite{Tagirov99} to be a prerequisite for a large spin valve effect. On the other hand, differences arise when non-collinear configurations are considered. In this case the dependence of $T_c$ on the angle $\alpha$ between the two magnetization directions is non-monotonic with a minimum for $\alpha=\pi/2$~\cite{Demler97}. A good agreement between theory and experiments is in this case obtained taking explicitly into account the odd triplet correlations generated via proximity effect by the non-collinearity of the magnetizations~\cite{Zhu10}.
Interestingly, it has been recently demonstrated that in F/S/F trilayers based on $d$-wave superconductors, such as in particular YBa$_2$Cu$_3$O$_{7-\delta}$ (YBCO) sandwiched between insulating layers of ferromagnetic Pr$_{0.8}$Ca$_{0.2}$MnO$_3$, the critical temperature shift between parallel and antiparallel configurations can approach the very large value of 2 K, with oscillations driven by the YBCO thickness over a length scale that is two orders of magnitude larger than the superconducting coherence length $\xi_\textsc{S}$~\cite{DiBernardo2019}.  
 
However, the F/S/F heterostructures so far considered have been theoretically investigated  by assuming for the ferromagnetic layers Stoner-like models where the bands associated with the two possible electron spin orientations have the same dispersion and are rigidly shifted in energy by the exchange interaction [Fig.~\ref{DOS}(a)]. Given the complexity of the forms in which the phenomenon of ferromagnetism manifests itself in metals, it may be relevant to perform the analysis of the above systems referring to scenarios different from the Stoner one. Among them, we will consider here a form of itinerant ferromagnetism driven by a gain in kinetic energy stemming from a spin-dependent bandwidth renormalization [Fig.~\ref{DOS}(b)], or, equivalently, from an effective spin mass mismatch (SMM) between spin-up and spin-down electrons~\cite{Hirsch99,Cuoco03}. 
Such kind of ferromagnetism can be theoretically described through microscopic approaches based on an extended Hubbard model, where the exchange and nearest-neighbor pair hopping terms are explicitly taken into account. Indeed, when treated within mean-field approaches, these contributions, generally neglected in studies based on the Hubbard model, lead to quasiparticle energies for the two spin species which are not simply split, as in the Stoner picture, but acquire different bandwidths or, equivalently, different effective masses~\cite{Hirsch}. For suitably low temperatures and in specific parameter regimes, this spin-dependent mass renormalization can lead to the establishment of a ferromagnetic order which arises from a gain in kinetic energy rather than in potential energy as in the usual Stoner scheme. 
This kind of ferromagnetism has experimentally been shown to be at the  origin of the optical properties of the colossal magnetoresistance in manganites~\cite{Okimoto}, in some rare-earth hexaborides~\cite{hexabor} as well as in some magnetic semiconductors~\cite{semicond}. As far as theory is concerned, this is predicted to substantially affect the coexistence of ferromagnetism and superconductivity~\cite{Ying03} as well as proximity~\cite{Cuoco08} and transport~\cite{Annunziata09,Annunziata11,Annunziata11b} phenomena in F/S bilayers. It may also play a role in the stabilization of the Fulde-Ferrell-Larkin-Ovchinnikov phase in heavy-fermion systems~\cite{HF}. Compared to the Stoner case, the interplay of this form of ferromagnetism with superconductivity is expected to give rise to different features in the behavior of several physical quantities. In particular, this issue has been investigated for F/S bilayer with anisotropic singlet superconductors, showing that the different response predicted for the two kinds of ferromagnets allows to discriminate among the possible time-reversal symmetry-breaking states established in the superconducting layer~\cite{Annunziata11}. In that context it was also shown that in a wide range of interface transparencies a SMM ferromagnet may support spin currents significantly larger than a standard Stoner one.

In this paper, the comparison between the role played by Stoner and SMM ferromagnets is performed referring to a clean F/S/F trilayer, in analogy with the study presented in Refs.~\cite{Annunziata09,Annunziata11} for a bilayer structure. In particular, by analyzing the behavior of the differential charge and spin conductances, we show that when the F/S/F junction is based on SMM ferromagnets, it behaves differently from junctions with ordinary Stoner ferromagnets, both in the transparent and in the tunnel limit. 
By comparing the transport properties of the junction in the superconducting regime (F/S/F) with those of the non-superconducting case (F/N/F), we find that for the SMM mechanism the interplay between superconductivity and ferromagnetism is not detrimental to charge and spin transport, as for the Stoner mechanism, but instead both charge and spin conductances through the F/S/F junctions get enhanced with respect to the F/N/F case in the bias region where superconductivity is most effective. 
Moreover, a distinctive feature of the SMM mechanism emerges in the spin transport at large bias, at intermediate and low barrier transparency: the sign of the spin current is opposite to the sign of the lead magnetization, while in the Stoner case this sign difference is not seen. We clarify the origin of this behavior, also showing that the presence of superconductivity is able to amplify the magnitude of the spin current at the gap edge both in the Stoner and in the SMM case.

Such results are presented in the paper as follows. In Section II we formulate the microscopic model based on the Bogoliubov-de Gennes (BdG) equations, written in each region of the junction. In this framework, we discuss how to solve the scattering problem in order to derive the probability coefficients associated with the relevant scattering processes. Then we explain how to use such coefficients to calculate the charge and spin conductance through the junction. The obtained results are discussed in Section III, in connection to the behavior of the scattering coefficients, both for fully transparent interfaces and in the tunnel limit. Finally, Section IV is devoted to the conclusions. 

Additional details about the applied formal procedure are provided in the Appendices. In Appendix A we report the expression of the wave functions for the injection processes which are not reported in the main text; Appendix B contains the derivation of the probability current conservation; in Appendix C we derive the spin-dependent charge conductance through the junction; Appendix D reports the behavior of the critical injection angles below which the different scattering processes are allowed; finally, Appendix E shows the derivation of a symmetry property characterizing the scattering amplitudes in the SMM case for transparent barriers and perpendicular injection.

\section{The model}

\begin{figure}[htbp]
	\centering
\includegraphics[width=0.98\columnwidth, angle=0]{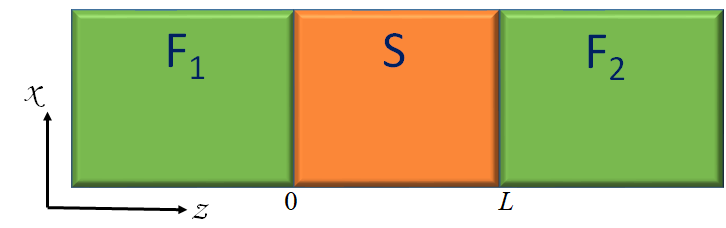}
\caption{Schematic representation of the considered symmetric planar ferromagnet/superconductor/ferromagnet (F/S/F) junction. $L$ is the thickness of the superconducting layer. The two ferromagnetic subsystems are assumed identical and semi-infinite.}
\label{fig1}
\end{figure}

We consider a planar symmetric trilayer junction in the clean limit, made up of a  superconducting layer of thickness $L$, sandwiched between two identical semi-infinite itinerant ferromagnets, as schematically shown in Fig.~\ref{fig1}. The junction lies in the $xz$-plane; the superconducting layer is connected to the two ferromagnetic electrodes by thin, insulating interfaces, located at the positions $z=0$ and $z=L$, respectively.

The barrier potential at the two interfaces is modelled as 
\begin{eqnarray}
V({\bf r})=H\delta(z)+H\delta(z-L)\, ,
\end{eqnarray}
\noindent where $H$ denotes the potential amplitude at each interface, and $\delta(z)$ is the Dirac delta function.
We also assume a rigid pairing potential for the superconducting side, such that
\begin{eqnarray}
\Delta_0({\bf r})=\Delta \Theta(z)\Theta(L-z)\, ,
\end{eqnarray}
\noindent where $\Theta(z)$ is the Heavyside step function and $\Delta_0$ is the BCS bulk gap.

Moreover, we assume that the effective mass of the system is 
\[
m^{\ast} ({\bf r}, \sigma) = m_{\sigma} \Theta(-z)  + m_{\sigma}  \Theta(z-L) + m_{\textsc{S}}\Theta(z)\Theta(L-z) \; .
\]
Here, $m_{\sigma}$ is the spin-dependent mass of electrons in each ferromagnetic layer, while $m_{\textsc{S}}$ is the carrier mass in the superconducting one. In order to analyze the effects on the transport across the junction which derive from the asymmetric mass renormalization, in comparison with those due to the conventional Stoner-induced ferromagnetism, we assume an equal exchange field in the two ferromagnetic sides of the junction: 
\[
h({\bf r}) = U \theta(-z) +U \theta(L-z) \; .
\]
We will illustrate in the following the solution of the quantum problem of the electron propagation from one side of the junction to the other, then showing how to determine the differential charge and spin conductances through the junction in the ballistic limit.

\subsection{Bogoliubov-de Gennes equations}

The single-particle Hamiltonian for a given spin projection $\sigma=\uparrow, \downarrow$ reads as
\begin{equation}
H^{\sigma}({\bf r})=-\frac{\hbar^{2}}{2\,m^{*}({\bf r}, \sigma)} \nabla^{2}\,+\, V({\bf r}) - \mu({\bf r}) - \rho_{\sigma} h({\bf r}) 
\label{BdGeqs}
\end{equation}
where $\rho_{{\uparrow}({\downarrow})}=+1(-1)$ and we have defined the chemical potential $\mu({\bf r})$ as
\begin{eqnarray}
\mu({\bf r}) & = &  E_{F}^{\textsc{F}} \ \theta(-z) + \ E_F^{\textsc{S}} \ \theta(z)\theta(L-z)+\ E_{F}^{\textsc{F}} \ \theta(z-L) \, , \nonumber \\
& &
\end{eqnarray}
with $E_{F}^{\textsc{F}}$ ($E_F^{\textsc{S}}$) being the Fermi energy in the ferromagnetic (superconducting) side.
We assume that there is no Fermi energy mismatch between the three subsystems, so that 
\begin{equation}
E_F\equiv E_{F}^{\textsc{F}}=E_F^{\textsc{S}} \; .
\label{Fermienergy}
\end{equation}
In the absence of spin-flip scattering, the spin-dependent four-component BdG equations for each subsystem can be decoupled into two subsets of two-component equations, one for the spin-up electron-like and spin-down hole-like quasiparticle wavefunctions $(u_{\uparrow},v_{\downarrow})$, and the other one for the corresponding quasi-particle wavefunctions having opposite spin projection $(u_{\downarrow},v_{\uparrow})$. The BdG equations for each subset are then:
\begin{eqnarray}
\left(
 \begin{array}{cc} 
 H^{\sigma}({\bf r}) &  \Delta_0({\bf r}) \\ 
  \Delta_0^{\ast}({\bf r}) &  -H^{\bar{\sigma}}({\bf r}) 
 \end{array} 
 \right)
\Psi_\sigma 
 =\varepsilon \Psi_\sigma \; , 
\label{HBdG}
\end{eqnarray} 
where $\bar{\sigma}=-\sigma$ and $\Psi_\sigma\equiv\left(u_\sigma,v_{\bar{\sigma}}\right)$
is the energy eigenstate in the electron-hole space associated with the eigenvalue $\varepsilon$.
The Hamiltonian invariance under translation along the $x$-direction allows to factorize the part of the eigenstates which corresponds to the electron motion in the direction parallel to the interfaces direction, i.e. $\Psi_\sigma(\mathbf{r})=e^{i\mathbf{k}_\parallel\cdot\mathbf{r}}\psi_\sigma(z)$, hence reducing the BdG problem to the solution of effective one-dimensional equations.
 
The solutions of the BdG equations for electrons ($e$) and holes ($h$) propagating in each ferromagnetic side are
\begin{eqnarray}
\psi_{\pm, \sigma}^{e}(z)= 
\left(
\begin{array}{c}
1 \\
0
\end{array} 
\right) e^{\pm \mathrm{i} q_{\sigma z}^{e} z}  \label{psie}\\
\psi_{\pm, \sigma}^{h}(z)= 
\left(
\begin{array}{c}
0 \\
1
\end{array}  
\right) e^{\pm \mathrm{i} q_{\sigma z}^{h} z} \label{psih}
\end{eqnarray}
where $q_{\sigma z}^{e(h)}$ is the projection along the $z$ direction of the electron (hole) momentum, whose total amplitude is
\begin{eqnarray}
q_{\sigma}^{e} & = &\sqrt{\frac{2\,m_{\sigma}}{\hbar^{2}}\left(E_{F}\,+\,\rho_{\sigma}\,U\,+\,\varepsilon\right)}\\
q_{\sigma}^{h} & = &\sqrt{\frac{2\,m_{\sigma}}{\hbar^{2}}\left(E_{F}\,+\,\rho_{\sigma}\,U\,-\,\varepsilon\right)} 
\end{eqnarray}
with $\rho_{\uparrow(\downarrow)}= +1\,(-1)$. The plus sign in Eq.(\ref{psie}) refers to electrons propagating from the left to the right side, while the minus sign indicates electrons moving in the opposite direction. Since holes have opposite group velocity direction with respect to electrons, in Eq.(\ref{psih}) the plus sign refers to hole motion from right to left while the minus one refers to hole propagation from left to right.

In the superconducting side, solutions for electron-like and hole-like quasiparticles are given by
\begin{eqnarray}
\psi_{\pm, \textsc{S}}^{e}(z) & = & 
\left(
\begin{array}{c}
u_0 \\
v_0
\end{array}
\right) e^{\pm i k_{z}^{e} z}
\label{psiSe}\\
\psi_{\pm, \textsc{S}}^{h}(z) & = &
\left(
\begin{array}{c}
v_0\\
u_0
\end{array}
\right) e^{\pm i k_{z}^{h} z} \; ,
\label{psiSh}
\end{eqnarray}
where $k_{z}^{e(h)}$ are the $z$ components of the electron-like (hole-like) quasiparticle momenta having amplitudes
\begin{eqnarray}
k^{e} & = &\sqrt{ \frac{2\,m_{\textsc{S}}}{\hbar^{2}}\left(E_{F}\,+ \sqrt{\varepsilon^{2}\,-\,\Delta^{2}}\right)}  \\
k^{h} & = &\sqrt{ \frac{2\,m_{\textsc{S}}}{\hbar^{2}}\left(E_{F}\,- \sqrt{\varepsilon^{2}\,-\,\Delta^{2}}\right)} \; ,
\end{eqnarray}
\noindent and $u_0$ and $v_0$ are the coherence factors expressed as 
\begin{eqnarray}
u_0 & = & \sqrt{\frac{1}{2}\left(1 + \frac{\sqrt{\varepsilon^{2} - \Delta^{2}}}{\varepsilon}\right)} \\
v_0 & = & \sqrt{ \frac{1}{2}\left(1 - \frac{\sqrt{\varepsilon^{2} - \Delta^{2}}}{\varepsilon}\right)}\; .
\end{eqnarray}
We apply the quasiclassical Andreev approximation, assuming that the processes of interest in our analysis involve quasiparticles which are close to the Fermi energy, such that $E_F \gg (\varepsilon, \Delta)$.
As a consequence, $q_{\sigma}^{e}=q_{\sigma}^{h} \equiv q_{\sigma}=\sqrt{\frac{2\,m_{\sigma}}{\hbar^{2}}\left(E_{F}\,+\,\rho_{\sigma}\,U\right)}$. 
We keep the energy dependence in the superconducting momenta occurring in the exponents of the superconducting wave functions given by Eqs.~(\ref{psiSe}) and (\ref{psiSh}), 
in such a way to catch the interference effects in the S region.
Using the Fermi energy condition (\ref{Fermienergy}) and the relation $E_F^{\textsc{S}}= \frac{\hbar^2 (k_F^\textsc{S})^2}{2 m_S}$, we can define the renormalized momentum amplitudes in the ferromagnetic and in the superconducting layers as
\begin{eqnarray}
\tilde{q}_{\sigma} &\equiv & \frac{q_{\sigma}}{k_F^{\textsc{S}}}=\sqrt{\sqrt{\frac{m_{\sigma}}{m_{\bar{\sigma}}}}\left(1+\rho_{\sigma} X \right)}
\label{q_sigma}\\
\tilde{k}^{e} & \equiv & \frac{k^{e}}{k_F^{\textsc{S}}}= \sqrt{1+\sqrt{\xi^2 - \delta^2}}\\
\tilde{k}^{h} & \equiv & \frac{k^{h}}{k_F^{\textsc{S}}}= \sqrt{1-\sqrt{\xi^2 - \delta^2}} \; ,
\end{eqnarray} 
respectively, having introduced the adimensional quantities 
$X=U/E_F$, $\xi=\varepsilon/E_F$, $\delta=\Delta/E_F$, with the condition $m_{\uparrow}/m_{\textsc{S}}=m_{\textsc{S}}/m_{\downarrow}$.

\subsection{The scattering problem}

The total wave function of the F/S/F trilayer is obtained as a linear combination of the solutions of the BdG equations for each individual region.
\begin{figure}[h!]
\centering
\includegraphics[width=0.95\columnwidth, angle=0]{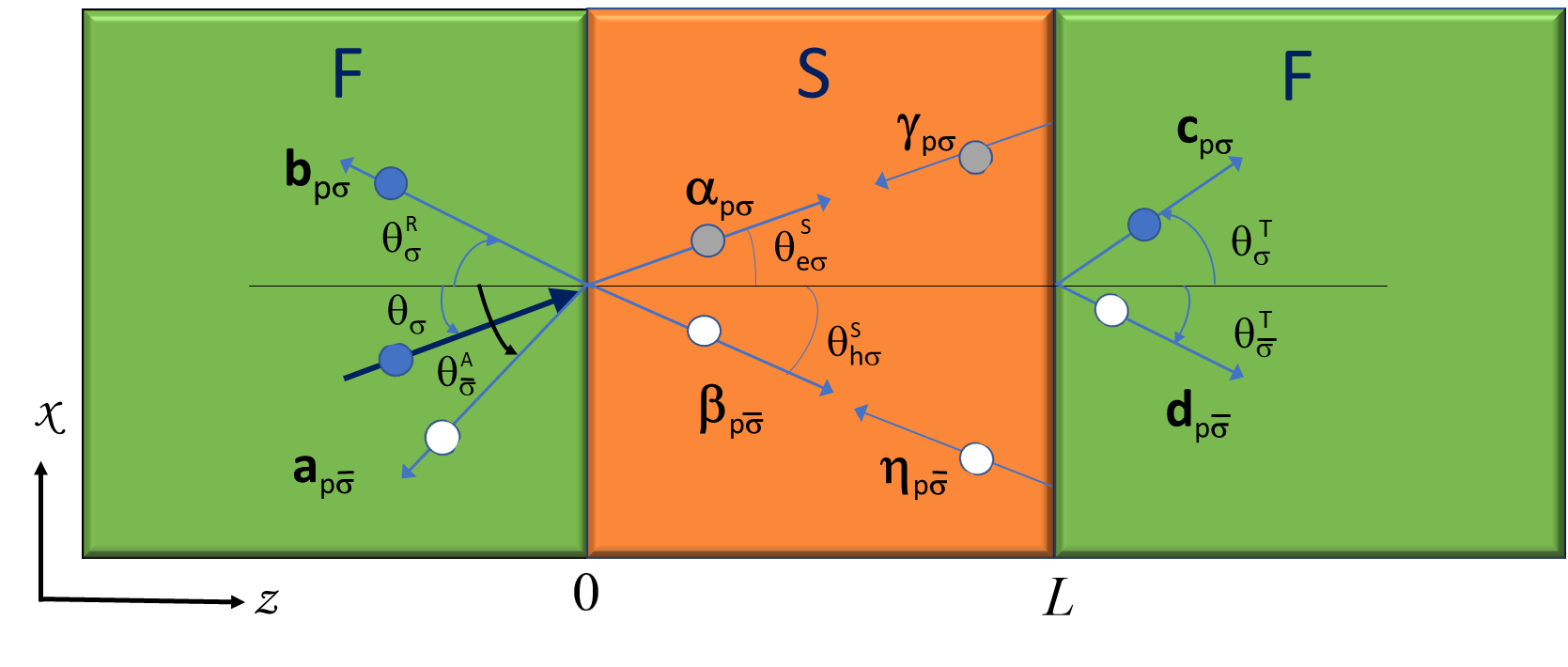}
\caption{Schematic representation of the scattering processes of a carrier ($\textsf{p} =e, h$ for electrons and holes, respectively) injected from the left ferromagnetic side with spin projection $\sigma$ at an angle $\theta_\sigma$ with respect to the direction perpendicular to the interfaces. The processes are described in detail in the text.}
\label{fig:scattering}
\end{figure}
Here we extend to the F/S/F trilayer the Blonder-Tinkham-Klapwjik (BTK) scattering theory~\cite{BTK} originally formulated for a N/S bilayer junction and then extended to a F/S one, in the case of a Stoner ferromagnet~\cite{Demler97} or a SMM one~\cite{Yoshida12}. This kind of approach has also been used to investigate a F/S/F junction, but only in the case of ferromagnetic order of the conventional Stoner type~\cite{Dong03,yamashita03,bosovic2003}. Rather, here the approach will also be applied to a trilayer with ferromagnetic layers of the SMM type.

When a carrier with spin $\sigma$ and momentum amplitude $q_{\sigma}$ is injected from the left ferromagnetic side with an injection angle $\theta_\sigma$, its propagation gives rise to eight possible scattering processes at the interfaces, as shown in Fig.~\ref{fig:scattering}. Within the left F layer they include $(i)$ Andreev reflection, converting the incident electron (hole) with spin $\sigma$ into a hole (electron) with opposite spin $\bar{\sigma}$, which propagates with a momentum amplitude $q_{\bar{\sigma}}$ along a direction forming an angle $\theta^{A}_{\bar{\sigma}}$ with the normal to the interface (\textsf{a}$_{\textsf{p}\bar{\sigma}}$ in Fig.~\ref{fig:scattering}), and $(ii)$ normal reflection as an electron (hole) having spin $\sigma$, momentum amplitude $q_{\sigma}$ and moving along a direction forming an angle $-\theta_{\sigma}$ with the normal direction to the interface (\textsf{b}$_{\textsf{p} \sigma}$ in Fig.~\ref{fig:scattering}). Then, in the S layer, close to the first interface, there occur $(i)$ the transmission of an electron in the superconducting side, propagating with momentum amplitude $k^{e}$ at an angle $\theta^{\textsc{S}}_{e \sigma}$ with respect to the normal to the interface ($\alpha_{\textsf{p} \sigma}$ in Fig.~\ref{fig:scattering}), and $(ii)$ the transmission of a hole, propagating with momentum amplitude $k^{h}$ and negative momentum component along the $z$-direction, at an angle $\theta^{\textsc{S}}_{h \sigma}$ with respect to the normal to the interface ($\beta_{\textsf{p} \bar{\sigma}}$ in Fig.~\ref{fig:scattering}). In proximity of the second barrier, the relevant processes are $(i)$ the reflection of an electron with momentum amplitude $k^{e}$ forming an angle $\theta^{\textsc{S}}_{e\sigma}$ with the normal direction to the interface ($\gamma_{\textsf{p}\sigma}$ in Fig.~\ref{fig:scattering}), and $(ii)$ the reflection of a hole having momentum amplitude $k^{h}$ and angle $\theta^{\textsc{S}}_{h\sigma}$ measured from the normal to the interface ($\eta_{\textsf{p}\bar{\sigma}}$ in Fig.~\ref{fig:scattering}). Finally, in the right F lead there occur $(i)$ the transmission of an electron with spin $\sigma$, momentum amplitude $q_{\sigma}$ along a direction forming an angle $\theta^{T}_{e\sigma}$ with the normal to the interface (\textsf{c}$_{\textsf{p} \sigma}$ in Fig.~\ref{fig:scattering}), and $(ii)$ the transmission of a hole with spin $\bar{\sigma}$, momentum amplitude ${q}_{\bar{\sigma}}$ along a direction forming an angle $\theta^{T}_{h \bar{\sigma}}$ with the normal to the interface (\textsf{d}$_{\textsf{p}\bar{\sigma}}$ in Fig.~\ref{fig:scattering}). 
Due to the symmetry of the junction, if quasiparticles are injected from the right F side, identical processes will occur, with reversed sign for all propagation velocities.

The scattering angles associated with the above listed processes depend on the injection angle $\theta_{\sigma}$ of the incident particles. They can be calculated by using the conservation of the momentum components which are parallel to the interfaces: 
\begin{eqnarray}
q_{\sigma \parallel} & \equiv & q_{\sigma} \sin \theta_{\sigma}=q_{\bar{\sigma}}  \sin \theta^{A}_{\bar{\sigma}} = k^{e(h)} \sin \theta^{\textsc{S}}_{e(h)\sigma} \nonumber \\
& = & q_{\sigma}  \sin \theta^{T}_{\sigma}=q_{\bar{\sigma}} \sin \theta^{T}_{\bar{\sigma}}\, . \label{BCangle2}
\end{eqnarray}
The scattering processes happening due to the injection of a quasiparticle $\textsf{p}=e,h$ from one of the two ferromagnetic sides in general occur with different probabilities, depending on the excitation energy $\varepsilon$, the superconducting energy gap $\Delta$, the polarization of the ferromagnetic leads, and the barrier strength $H$. Consequently, the scattering processes enter the total wave function expression with some unknown amplitudes to be determined by imposing the matching conditions for the wave functions at the interfaces. 

Taking into consideration the above listed  scattering processes, the wave function for an electron which is injected from the left F side at energy $\varepsilon$, angle $\theta_{\sigma}$ and with spin $\sigma$ can be written as 
\begin{eqnarray}
\nonumber \psi_{e\sigma \textsf{L}}^{\textsc{F}}(z)&=&
\left(
\begin{array}{l}
 1\\
 0 
\end{array}
\right) e^{\mathrm{i} q_{\sigma} z \ cos\theta_{\sigma}}
+  \textsf{a}_{\textsf{e}\bar{\sigma}}  
\left(
\begin{array}{l}
0\\
1 
\end{array}
\right)
e^{\mathrm{i} q_{\bar{\sigma}} z \ cos\theta^{A}_{\bar{\sigma}}} \\
&& + \, \textsf{b}_{\textsf{e} \sigma}   
\left(
\begin{array}{l}
1\\
0 
\end{array}
\right)
e^{- \mathrm{i} q_{\sigma} z \ cos\theta_{\sigma}} \label{psi1E}
\end{eqnarray}
in the left F side ($z<0$), as 
\begin{eqnarray}
\nonumber  \psi^{\textsc{S}}_{e \sigma} (z)&=& 
\alpha_{\textsf{e} \sigma }  
\left(
\begin{array}{l}
 u_0\\
 v_0 
\end{array}
\right) 
e^{\mathrm{i} k^{e} z \ cos\theta_{\sigma e}^{\textsc{S}}}
\\ 
\nonumber & & + \, \beta_{\textsf{e}\bar{\sigma}}   
\left(
\begin{array}{l}
v_0\\
u_0 
\end{array}
\right)
e^{- \mathrm{i} k^{h} z \ cos\theta_{\sigma h}^{\textsc{S}}}
\\ 
\nonumber & & + \,
\gamma_{\textsf{e} \sigma }   \ 
\left(
\begin{array}{l}
 u_0\\
 v_0 
\end{array}
\right) 
e^{-\mathrm{i} k^{e} z \ cos\theta_{\sigma e}^{\textsc{S}}}
\\ 
& & + \, \eta_{\textsf{e}\bar{\sigma}}    
\left(
\begin{array}{l}
v_0\\
u_0 
\end{array}
\right)
e^{\mathrm{i} k^h z \  cos\theta_{\sigma h}^{\textsc{S}}}  \label{psiSE}
\end{eqnarray}
in the S side ($0<z<L$), and as
\begin{eqnarray} 
\nonumber \psi_{e\sigma \textsf{R}}^{\textsc{F}}(z)&=&
\textsf{c}_{\textsf{e}\sigma} 
\left(
\begin{array}{l}
1\\
0
\end{array}
\right)
e^{\mathrm{i}q_{\sigma}z \ cos\theta^{T}_{\sigma}}+
\\&&
\textsf{d}_{\textsf{e}\bar{\sigma}} \left(
\begin{array}{l}
0\\
1 
\end{array}
\right)
 e^{-\mathrm{i}q_{\bar{\sigma}}z \ cos\theta^{T}_{\bar{\sigma}}} \label{psi2E}
\end{eqnarray} 
in the right F side ($z>L$). The corresponding wave functions for the injection of a hole from the left F lead are reported in Appendix A. 

The coefficients corresponding to the scattering processes occurring at the two interfaces are determined by imposing the following boundary conditions: the continuity conditions of the wave functions at the two interfaces,
\begin{eqnarray}
\psi^{\textsc{F}}_{p\sigma \textsf{L}} \left(0\right) & = & \psi^{\textsc{S}}_{p\sigma}\left(0\right) \nonumber \\
\psi^{\textsc{F}}_{p\sigma \textsf{R}}\left(L\right) & = & \psi^{\textsc{S}}_{p\sigma}\left(L \right) \; ,
\label{contpsi}
\end{eqnarray} 
and the discontinuity of the wave function first derivative with respect to the $z$ spatial coordinate at the interface locations due to the local barrier potentials. Such conditions are derived by integrating the BdG equations in Eqs.~(\ref{HBdG}) over the $z$ variable in a very narrow range around each barrier, and read:
\begin{eqnarray}
\frac{d}{d z}u^{\textsc{S}}_{\sigma}\Big \vert _{z = 0} - \frac{m_{\textsc{S}}}{  m_{\sigma}}  \frac{d}{d z}u^{\textsc{F}}_{\sigma \textsf{L}}\Big \vert _{z = 0} & = & \textsf{Z} u^{\textsc{S}}_{\sigma}\left(0\right)  \nonumber \\
  \frac{d}{d z} v^{\textsc{S}}_{\bar{\sigma}}\Big \vert _{z = 0} - \frac{m_{\textsc{S}}}{m_{\bar{\sigma}}}  \frac{d}{d z}v^{\textsc{F}}_{\bar{\sigma} \textsf{L}}\Big \vert _{z = 0} & = & \textsf{Z} v^{\textsc{S}}_{\bar{\sigma}}\left(0\right) \nonumber \\
 \frac{m_{\textsc{S}}}{m_{\sigma}}  \frac{d}{d z}u^{\textsc{F}}_{\sigma \textsf{R}}\Big \vert _{z = L} -  \frac{d}{d z} u^{\textsc{S}}_{\sigma}\Big \vert_{z = L} & = & \textsf{Z} u^{\textsc{S}}_{\sigma}\left(L\right) \nonumber \\
 \frac{m_{\textsc{S}}}{m_{\bar{\sigma}}}  \frac{d}{d z} v^{\textsc{F}}_{\bar{\sigma} \textsf{R}}\Big \vert _{z = L} - \frac{d}{d z} v^{\textsc{S}}_{\bar{\sigma}} \Big \vert _{z = L} & = & \textsf{Z} v^{\textsc{S}}_{\bar{\sigma}}\left(L\right)\; .
\label{BC_system}
\end{eqnarray}
Here $(u^{\textsc{S}}_{\sigma},v^{\textsc{S}}_{\bar{\sigma}})$ are the components of the superconducting wave function $\psi^{\textsc{S}}_{e \sigma} (z)$, whose explicit expression is given in (\ref{psiSE}), whereas $(u^{\textsc{F}}_{\sigma \alpha},v^{\textsc{F}}_{\bar{\sigma} \alpha})$ with $\alpha= \textsf{L}, \textsf{R}$  are the components of the ferromagnetic wave functions $\psi^{\textsc{F}}_{e \sigma \textsf{L} (\textsf{R})} (z)$ in the left ($\textsf{L}$) and right ($\textsf{R}$) side, respectively. Their explicit expressions have been provided in Eqs.~(\ref{psi1E}),(\ref{psi2E}). Finally, we have defined  $\textsf{Z}= \frac{2 m_{\textsc{S}} H}{\hbar^2 k_F^\textsc{S}}$. 

The probability amplitudes associated with each scattering process are obtained from the solution of the system in Eqs.~(\ref{BC_system}). This is done by using the conservation of the current probability, in the form derived in Appendix B. For a carrier $\textsf{p}=e,h$ injected from left with energy $\varepsilon=\tilde{\xi} \Delta$ along a direction forming an angle $\theta_{\sigma}$ with the direction normal to the interface, we find that the probabilities $\textsf{A}_{\textsf{p}\sigma}$ for local Andreev reflections, $\textsf{B}_{\textsf{p}\sigma}$ for the specular reflection, $\textsf{C}_{\textsf{p}\sigma}$ for the transmission with the same charge in the right F lead and $\textsf{D}_{\textsf{p}\sigma}$ for the transmission with opposite charge in the right F lead are given by:
\begin{eqnarray}
\textsf{A}_{\textsf{p}\sigma}& = &\frac{m_{\sigma}}{m_{\bar{\sigma}}} \frac{q_{\bar{\sigma}}\cos \theta^{A}_{\bar{\sigma}}}{q_{\sigma} \cos \theta_{\sigma}} \vert \textsf{a}_{\textsf{p}\bar{\sigma}}\left(\tilde{\xi}, \theta_{\sigma}\right)\vert^{2}\, ,
\label{eqApr}\\
\textsf{B}_{\textsf{p}\sigma} & = &\vert \textsf{b}_{\textsf{p}\sigma}\left(\tilde{\xi}, \theta_{\sigma}\right)\vert^{2}\,
\label{eqBpr} ,\\
\textsf{C}_{\textsf{p}\sigma} & = & \frac{ \cos \theta^{T}_{\sigma}}{\cos \theta_{\sigma}} \vert \textsf{c}_{\textsf{p}\sigma}\left(\tilde{\xi}, \theta_{\sigma}\right)\vert^{2} \, 
\label{eqCpr},\\
\textsf{D}_{\textsf{p}\sigma}& = &\frac{m_{\sigma}}{m_{\bar{\sigma}}}  \frac{q_{\bar{\sigma}}\cos \theta^{T}_{\bar{\sigma}}}{q_{\sigma}\cos \theta_{\sigma}} \vert \textsf{d}_{\textsf{p}\bar{\sigma}}\left(\tilde{\xi}, \theta_{\sigma}\right)\vert^{2}\,
\label{eqDpr} .
\end{eqnarray}
In the case of injection from the right side, the expressions are exactly the same as Eqs.~(\ref{eqApr})-(\ref{eqDpr}), due to the mirror symmetry of the junction.

\subsection{Charge and spin conductances}

\begin{figure}[h!]
 	\centering
  	\includegraphics[scale=0.4]{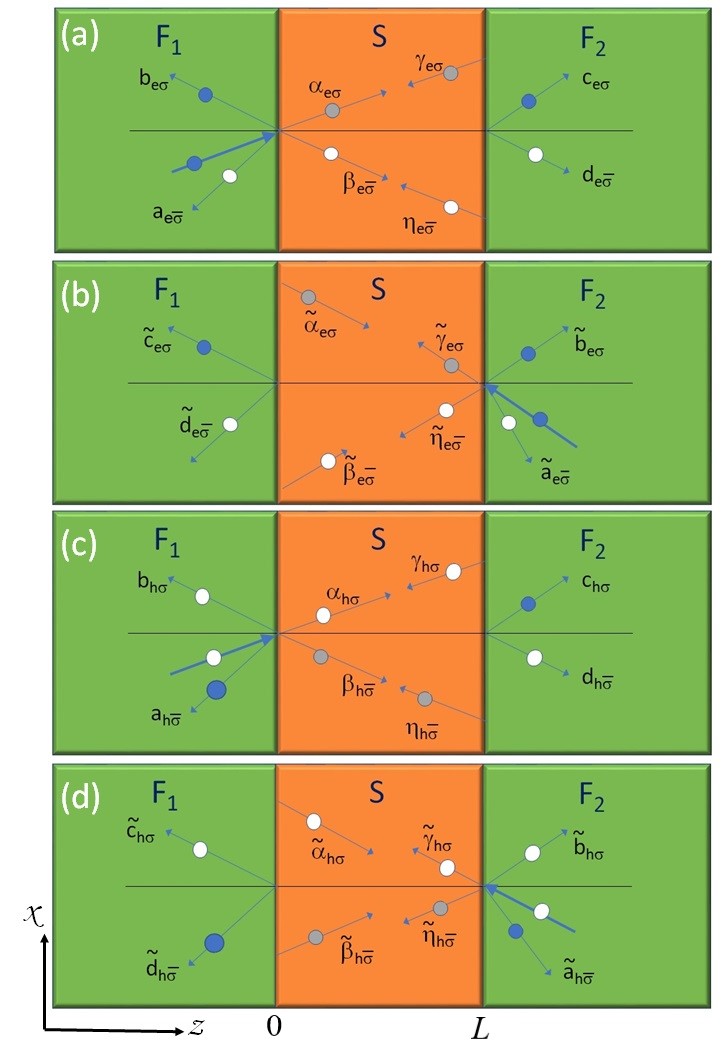}
  	\caption{Processes involved in the calculation of the charge and spin conductance through the junction: (a) injection of an electron with spin $\sigma$ from the left side of the junction; (b) injection of a hole with spin $\sigma$  from the left side; (c) injection of an electron with spin $\sigma$  from the right side, (d) injection of a hole with spin $\sigma$ from the right side. }\label{figprocessiok}
  \end{figure}

The knowledge of the coefficients (\ref{eqApr})-(\ref{eqDpr}) allows to obtain the expression of the charge and the spin conductances, again referring to the extension of the BTK approach to the case of F/S/F trilayer.
As for the case of the single F/S junction~\cite{Yoshida12}, the conductance can be more conveniently calculated into the left ferromagnetic side of the junction, where the current flow does not include supercurrents. In the presence of an applied bias $V$ between the left and right side of the junction, four injection processes can occur, as graphically shown in Fig.~\ref{figprocessiok}: the injection of an electron with spin $\sigma$ from the left side of the junction [Fig.~\ref{figprocessiok}(a)], the injection of a hole with spin $\sigma$ from the left side [Fig.~\ref{figprocessiok}(b)], the injection of an electron with spin $\sigma$  from the right side [Fig.~\ref{figprocessiok}(c)], and the injection of a hole with spin $\sigma$ from the right side [Fig.~\ref{figprocessiok}(d)]. 
As derived in detail in Appendix C, by properly taking into account all these processes ~\cite{yamashita03}, the charge and spin currents, $J_{c}$ and $J_s$ respectively, flowing normally to the interfaces due to the application of a voltage bias $V$, can be written as
\begin{equation}
J_{c(s)}=J_{\uparrow}\pm J_{\downarrow}
\label{Eq_ch&sp}
\end{equation}
where the spin-dependent current $J_{\sigma}$ is 
\begin{eqnarray}
\nonumber J_{\sigma} & = & e  \hbar \sum_{\epsilon, \theta \in F_1}  \frac{q_{\sigma}}{m_{\sigma}}\cos\theta \left[f\left(\epsilon - \frac {e V}{2}\right) - f\left(\epsilon + \frac {e V}{2}\right) \right] \\
&& \times \left(\textsf{A}_{e \sigma}  +  \textsf{C}_{e \sigma} + \textsf{A}_{h \sigma} +  \textsf{C}_{h\sigma} \right) \; .
\end{eqnarray}
Starting from this expression, one can show (see Appendix C) that the spin-dependent angle-averaged differential conductance for an applied bias $V$ can be written as 
\begin{equation}
G_\sigma(\textsf{E})=\frac{dJ_{\sigma}}{dV} = \int_{0}^{\theta^{c}_\sigma}
d\theta \ \textsc{G}_\sigma(\textsf{E},\theta)
\label{Eq_gsigma}
\end{equation}
with $\textsf{E}=e V$ and
\begin{equation}
\textsc{G}_{\sigma}(\textsf{E},\theta)= G_0 \ \tilde{q}_{\sigma} \cos\theta \left(\textsf{A}_{e \sigma}+ \textsf{C}_{e \sigma} + \textsf{A}_{h \sigma}+ \textsf{C}_{h\sigma}\right)\left|_{\frac{\textsf{E}}{2},\theta}\right.  \, .
\label{GthetaE}
\end{equation}
Here $G_0=\frac{e^2  q_F }{\pi \hbar} $ is the conductance of the junction when the three layers are all in the normal state, $q_F$ is the Fermi momentum in the normal state, and $\tilde{q}_{\sigma}= q_{\sigma}/ q_{F}$. 
The definition (\ref{Eq_gsigma}) of $G_\sigma$ takes into account that the experimentally measured conductance takes contributions from a limited range of injection angles, depending on the experimental conditions. This is specified by the value of $\theta^{c}_\sigma$, which is the critical incidence angle for electrons with spin $\sigma$ injected from the left ferromagnet, above which transmission processes to the right ferromagnet do not occur. An explicit evaluation of the critical angles characterizing the scattering processes taking place within the junction is presented in Appendix D.

In terms of $G_\sigma$ the charge and the spin conductance are defined as
\begin{eqnarray}
G_c(\textsf{E}) & = & G_\uparrow(\textsf{E}) + G_\downarrow(\textsf{E}) \\
G_s(\textsf{E}) & = & G_\uparrow(\textsf{E}) - G_\downarrow(\textsf{E}) \; .
\end{eqnarray}

\section{Results and discussion}

\begin{figure}[h!]
\centering
\includegraphics[width=0.98\columnwidth,  angle=0]{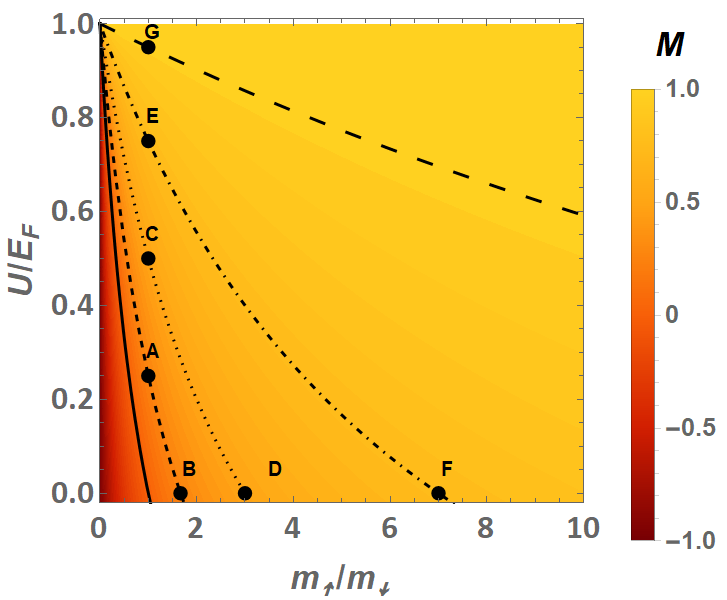}
\caption{Density plot of the magnetization for a two-dimensional ferromagnetic system where both Stoner and SMM mechanisms are responsible for the ferromagnetic state. Isomagnetic curves are shown: continuous, small dashed, dotted, dotted-dashed and large dashed lines correspond to $M=0, 0.25, 0.5, 0.75, 0.95$, respectively. The parameter values associated with each marked point are given in Table \ref{TabMag}.  Point H lies far away on the horizontal axis and cannot be properly shown in the figure.}
\label{figMAG}
\end{figure}

\begin{figure}[h!]
\centering
 \includegraphics[width=0.7\columnwidth, angle=0]{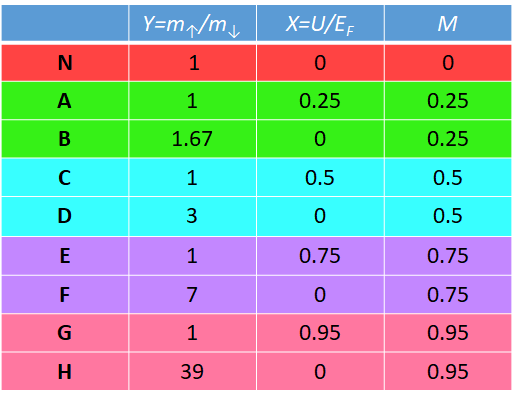}
\caption{Chosen values of the magnetization and corresponding microscopic parameters used to investigate separately the pure Stoner case and the SMM one.}
\label{TabMag}
\end{figure}

We assume a superconducting layer having thickness $L=5000/k_F$, which is of the order of the superconducting coherence lenght $\xi_{\textsc{S}}=2000/k_F$. 
In order to compare the effects due to the two microscopic mechanisms responsible for the ferromagnetism, we fix the magnetization amplitude in the F leads, and then we choose pairs of states where the same value of magnetization is obtained either via a pure Stoner-like mechanism or via the mass splitting one only. Referring to Fig.~\ref{figMAG}, showing the magnetization as a function of $X$ and of the mass ratio $Y=m_{\uparrow}/m_{\downarrow}$, a given pair of such states is represented by two points lying on the same isomagnetic curves (small dashed, dotted, dotted-dashed and large dashed lines for $M=0.25, 0.5, 0.75, 0.95$, respectively). Points such as A, C, E, and G correspond to $m_\uparrow=m_\downarrow$ and thus are representative of ferromagnetic states of pure Stoner origin, while points such as B, D, F along the horizontal axis correspond to pure SMM ferromagnetic states. The different cases analyzed in the following correspond to the values of $X$, $Y$ and $M$ reported in the table of Fig.~\ref{TabMag}.

By using Eqs.~(\ref{Eq_ch&sp})-(\ref{GthetaE}), we have calculated the charge and the spin conductance through the double junction at different values of the applied bias, of the polarization of the ferromagnetic leads, and of the barrier transparencies.
The charge and spin conductances of the F/S/F junction have both been normalized to the charge conductance of the corresponding ferromagnetic/normal/ferromagnetic (F/N/F) junction, in order to better visualize the effects induced by the superconducting state.

\subsubsection{Charge conductance}

\begin{figure}[htbp]
\includegraphics[width=0.48\columnwidth, angle=0]{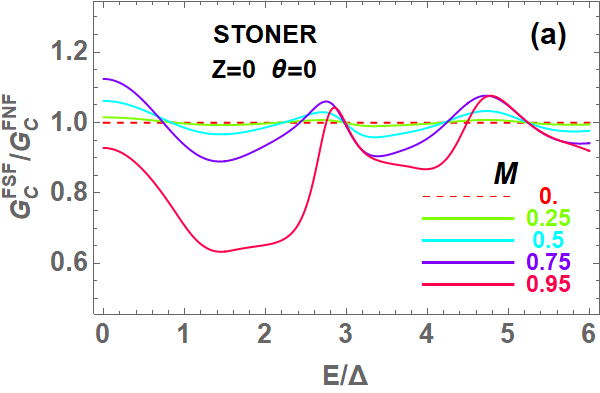}
\includegraphics[width=0.48\columnwidth, angle=0]{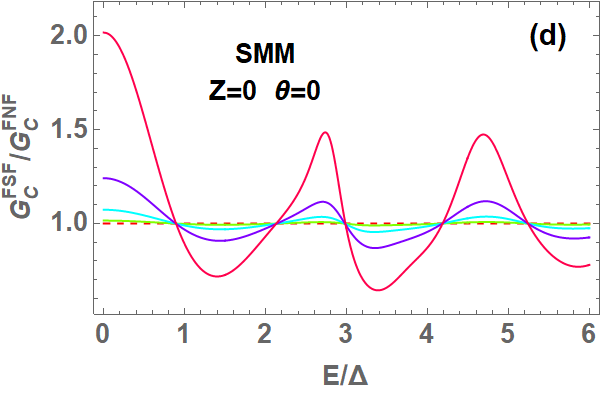}
\includegraphics[width=0.48\columnwidth, angle=0]{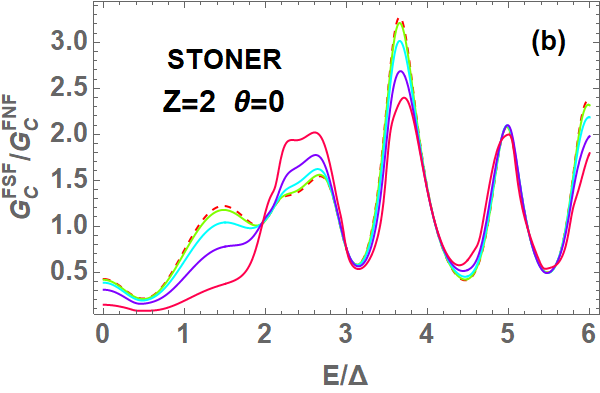}
\includegraphics[width=0.48\columnwidth, angle=0]{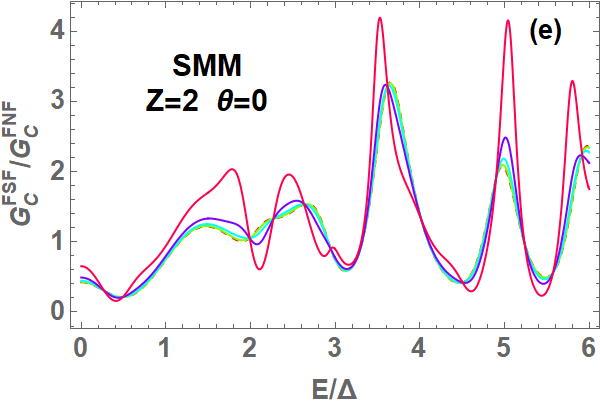}
\includegraphics[width=0.48\columnwidth, angle=0]{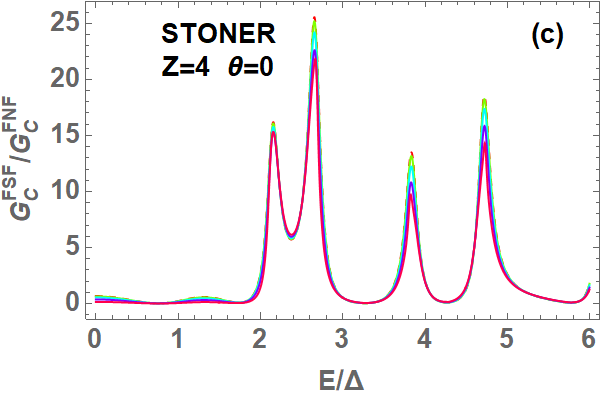}
\includegraphics[width=0.48\columnwidth, angle=0]{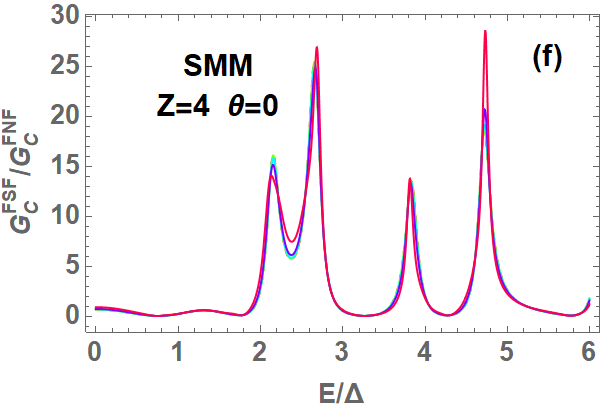}
\caption{Voltage bias ($\textsf{E}=e V$) dependence of the charge conductance at normal incident angle $(\theta=0)$ in the Stoner case [(a)-(c)] and in the SMM case [(d)-(f)] at different values of the barrier transparency $\textsf{Z}$: (a) and (d) refer to $\textsf{Z}=0$, (b) and (e) to $\textsf{Z}=2$, (c) and (f) to $\textsf{Z}=4$. The chosen values of the magnetization are $M=0, 0.25, 0.5, 0.75, 0.95$ (red, green, cyan, violet and magenta lines, respectively).}  
\label{Fig_condC_th0}
\end{figure}
  
The charge conductance of the F/S/F junction for particles injected perpendicularly to the barriers is shown in Fig.~\ref{Fig_condC_th0} in the Stoner case [Figs.~\ref{Fig_condC_th0}(a)-\ref{Fig_condC_th0}(c)] and in the SMM case [Figs.~\ref{Fig_condC_th0}(d)-\ref{Fig_condC_th0}(f)], for three different values of the barrier transparency. 
The most appreciable differences between the junction behavior in the presence of the two different mechanisms for ferromagnetism appear for fully transparent interfaces ($\textsf{Z}=0$) and high values of the magnetization. Indeed, in this regime, while in the Stoner case the charge conductance is significantly suppressed at low bias with respect to the F/N/F case, on the contrary in the SMM case it is enhanced when the value of the mass mismatch is increased.

This effect can be understood by taking into account the expression of $\textsc{G}_{\sigma}(\textsf{E},\theta)$, which explicitly depends on the carrier linear momentum [Eq.(\ref{GthetaE})]. Since a large magnetization directly affects the linear momentum of carriers involved in the transmission processes [see Eq.(\ref{q_sigma})] in a more sizable way in the SMM case than in the Stoner one, a large mass mismatch, such as the one occurring for the parameter choice corresponding to the point H listed in the table of Fig.~\ref{TabMag}, is expected to induce a strong contribution to the charge conductance. Correspondingly, at lower values of the magnetization, where the mass mismatch in SMM ferromagnets is reduced, carrier momenta for spin-up and spin-down electrons assumes more comparable values, which makes the SMM charge conductance more similar to the Stoner one.      

\begin{figure}[htbp]
\includegraphics[width=0.49\columnwidth, angle=0]{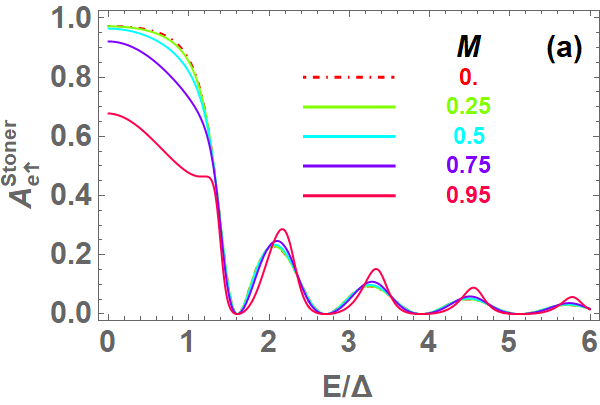}
 \includegraphics[width=0.49\columnwidth, angle=0]{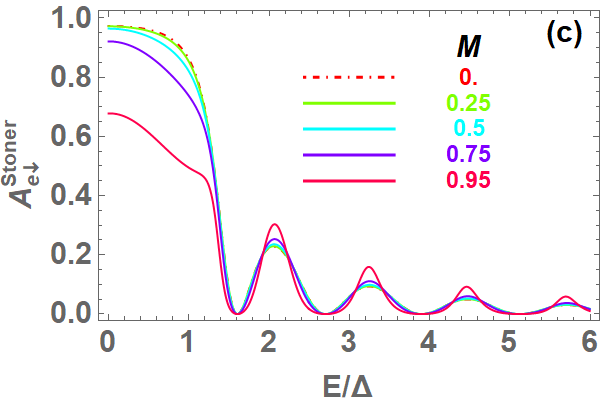}
\includegraphics[width=0.49\columnwidth, angle=0]{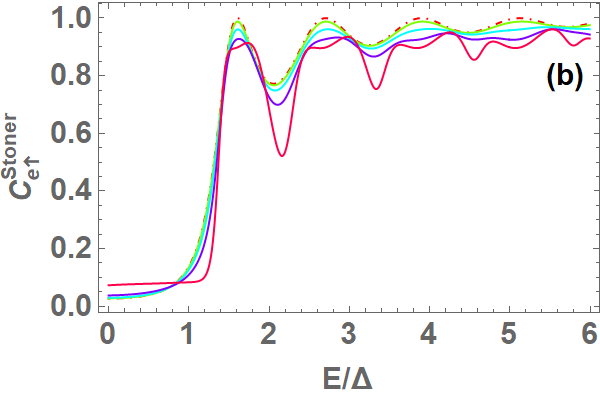}
\includegraphics[width=0.49\columnwidth, angle=0]{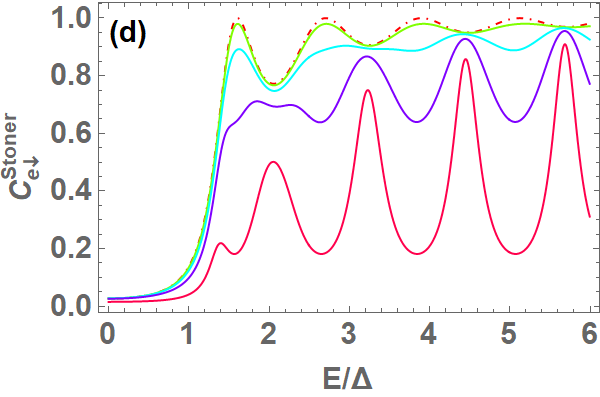}
\caption{Energy dependence of the probability coefficients for Andreev reflections ($\textsf{A}$) and transmission in the right ferromagnet as electrons ($\textsf{C}$), for spin-up and spin-down injected carriers [(a),(b) and (c), (d), respectively]. Here we have considered the case of Stoner ferromagnetic layers and high-transparent barriers ($\textsf{Z}=0)$. The values of $M$ and the related color lines are the same as those used in Fig.~\ref{Fig_condC_th0}.}
\label{ST1}
 \end{figure}

\begin{figure}[htbp]
\includegraphics[width=0.48\columnwidth, angle=0]{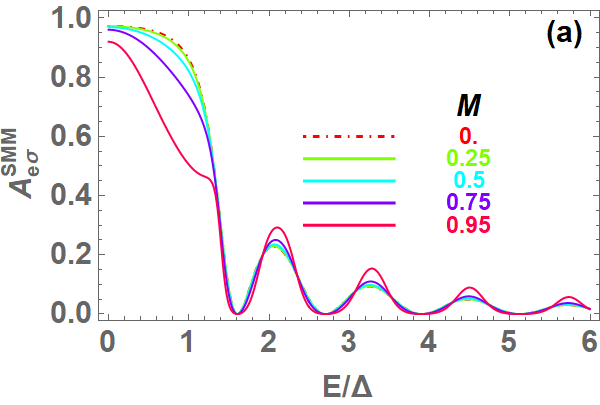}
\includegraphics[width=0.48\columnwidth, angle=0]{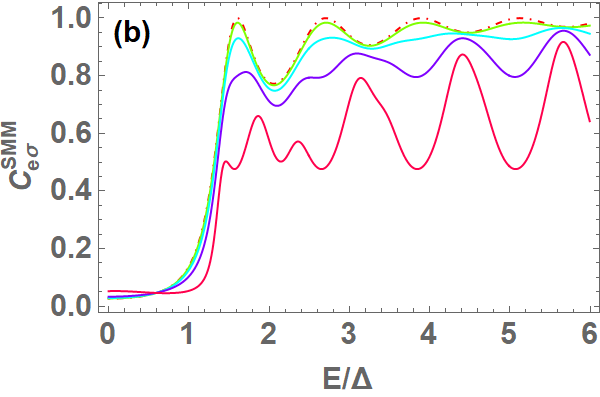}
\caption{Energy dependence of the probability coefficients for (a) Andreev reflections, (b) transmission into the right ferromagnet as electrons for spin-up injected carriers, in the case of SMM ferromagnetic layers and for perfectly transparent barriers ($\textsf{Z}=0$). The same coefficients for spin-down electrons are not shown because, as proved in Appendix E, they coincide with those for spin-up ones. The values of $M$ and the related color lines are the same as those used in Fig.~\ref{Fig_condC_th0}.}
\label{SBA1}
\end{figure}

Aside from the magnetization driven enhancement of the particle linear momentum, another effect generally contributes to determine the observed differences between the SMM and the Stoner charge conductance, and it is linked to the Andreev reflections which occur when superconductivity is switched on. They directly contribute to the conductance according to Eq.(\ref{GthetaE}), and, in particular, they play the major role in the scattering process for applied bias below the energy gap at large barrier transparencies [see Figs.\ref{ST1}\,(a), \ref{ST1}\,(c), and Fig.\ref{SBA1}\,(a)].  
The occurrence of Andreev scattering due to the superconducting state counterbalances the detrimental effect that ferromagnetism has on the charge conduction. Indeed, both mechanisms responsible for ferromagnetism generally disadvantage the charge transport through the junction. The Stoner mechanism, via the energy shift between the opposite spin electrons [Fig.\ref{DOS}(a)], reduces the available states for minority carriers; the SMM mechanism induces an unbalance between the velocities of carriers with opposite spin, such that at positive magnetization values, spin-up electrons become slower than spin-down ones, thus providing a reduced contribution to the conductance. Therefore, in the presence of superconductivity, the lack of the energy shift between opposite spin density of states in the SMM case allows strong Andreev reflections, which are instead suppressed by increasing the magnetization in the Stoner case, due to a reduction of accessible states for spin-down holes. Therefore, the more robust Andreev reflections, together with a sizable linear momentum amplification at large magnetization in the SMM case, can explain the very different behavior of the charge conductance of the F/S/F junctions in the Stoner and in the SMM case for transparent barriers [Fig.~\ref{Fig_condC_th0}(a) and \ref{Fig_condC_th0}(d)].


\begin{figure}[htbp]
\includegraphics[width=0.49\columnwidth, angle=0]{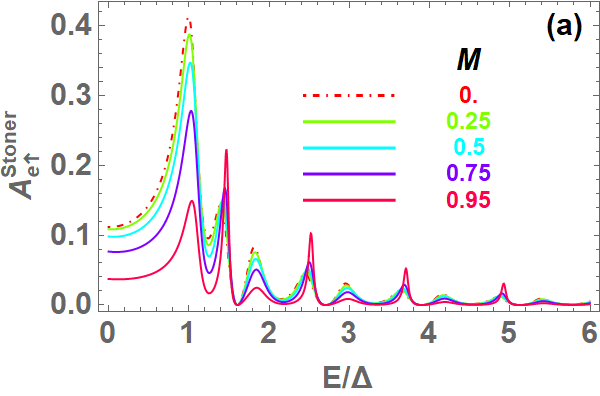}
\includegraphics[width=0.49\columnwidth, angle=0]{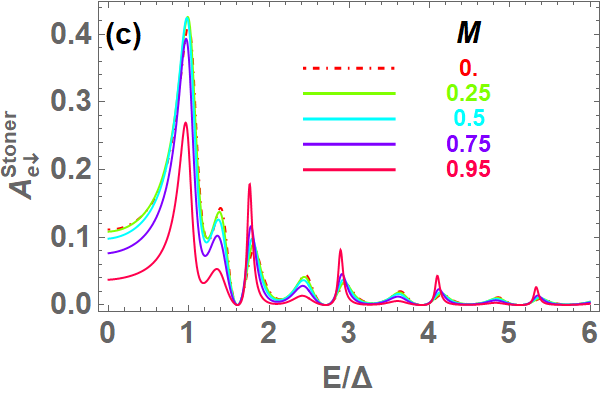}
\includegraphics[width=0.49\columnwidth, angle=0]{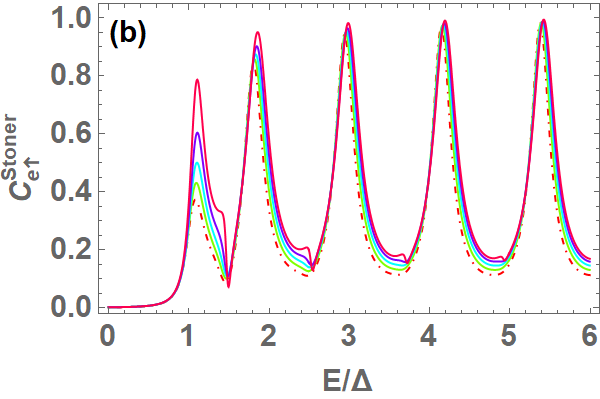}
\includegraphics[width=0.49\columnwidth, angle=0]{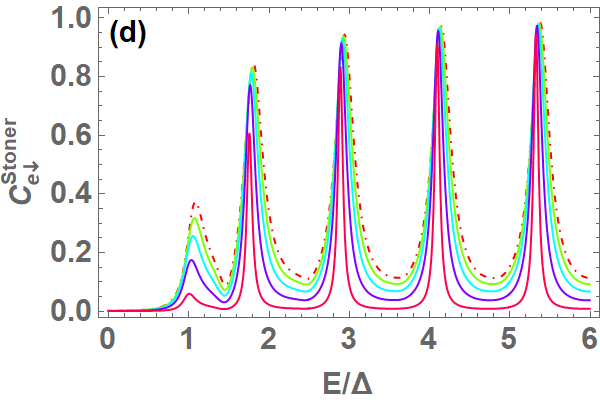}
\caption{Energy dependence of the probability coefficients in the Stoner case for Andreev reflections ($\textsf{A}$) and transmission to the right ferromagnet as electrons ($\textsf{C}$), for spin-up injected electrons [respectively (a) and (b)] and spin-down injected electrons [respectively (c) and (d)], in the low-transparency limit ($\textsf{Z}=2$). The values of $M$ and the related color lines are the same as those used in Fig.~\ref{Fig_condC_th0}.}
\label{CoeffSTZ2}
\end{figure}

\begin{figure}[htbp]
\includegraphics[width=0.49\columnwidth, angle=0]{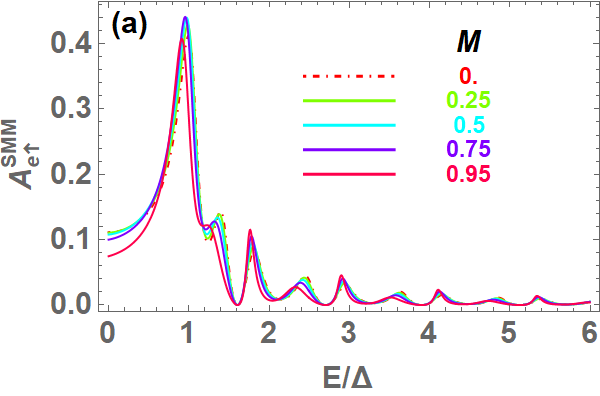}
\includegraphics[width=0.49\columnwidth, angle=0]{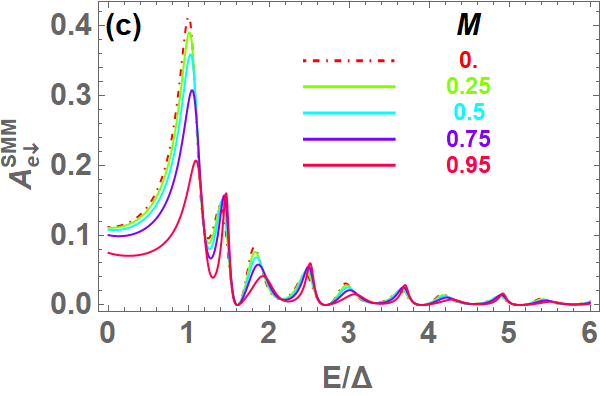}
\includegraphics[width=0.49\columnwidth, angle=0]{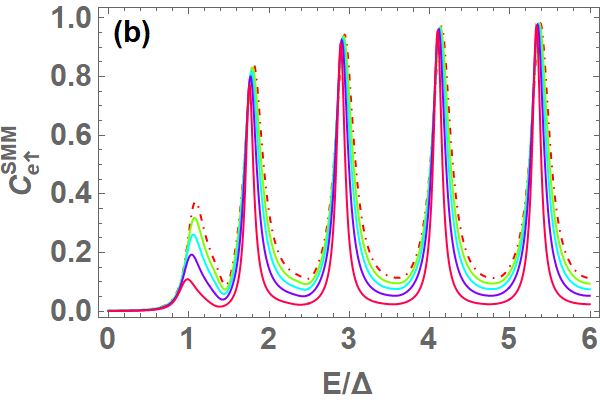}
\includegraphics[width=0.49\columnwidth, angle=0]{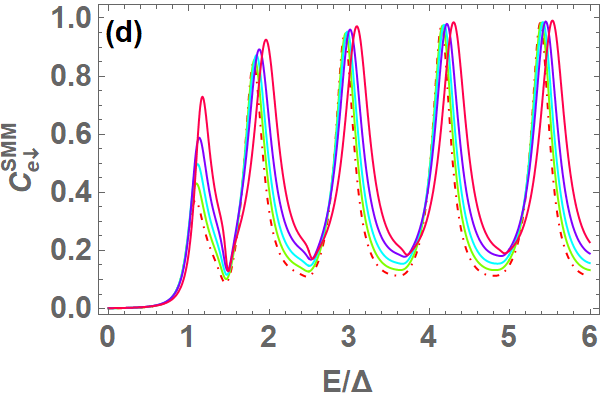}
   \caption{Energy dependence of the probability coefficients in the SMM case for Andreev reflections (A) and transmission to the right ferromagnet as electrons (C), for spin-up injected electrons [respectively (a) and (b)] and spin-down injected electrons [respectively (c) and (d)], in the low-transparency limit ($\textsf{Z}=2$). The values of $M$ and the related color lines are the same as those used in Fig.~\ref{Fig_condC_th0}.}
 \label{CoeffSBZ2}
\end{figure}

Differently from what happens in the regime of highly transparent barriers, for finite  transparency differences between the Stoner and the SMM case tend to become less and less appreciable as $\textsf{Z}$ is increased, regardless of the magnetization value in the F layers. This can be explained in terms of the behavior of the Andreev reflections, which at low bias become strongly suppressed by increasing $\textsf{Z}$ [see for instance Figs.~\ref{CoeffSTZ2} and \ref{CoeffSBZ2} for the case $\textsf{Z}=2$]. The small contribution provided by the Andreev reflections makes less and less effective the role played by the large momentum values associated with a high value of the mass mismatch, thus leading to a similar behavior of the Stoner and the SMM charge conductance at any value of $M$.   

\begin{figure}[htbp]
\includegraphics[width=0.45\columnwidth, angle=0]{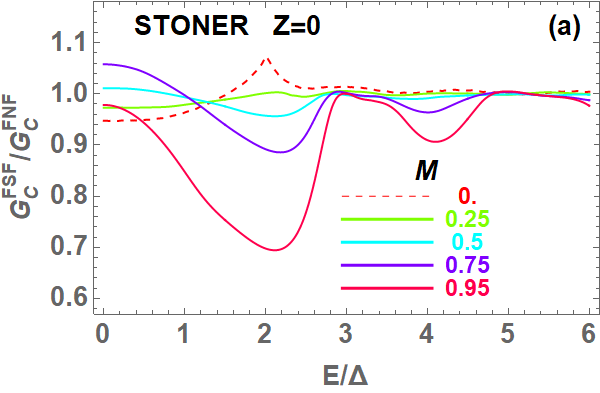}
\includegraphics[width=0.45\columnwidth, angle=0]{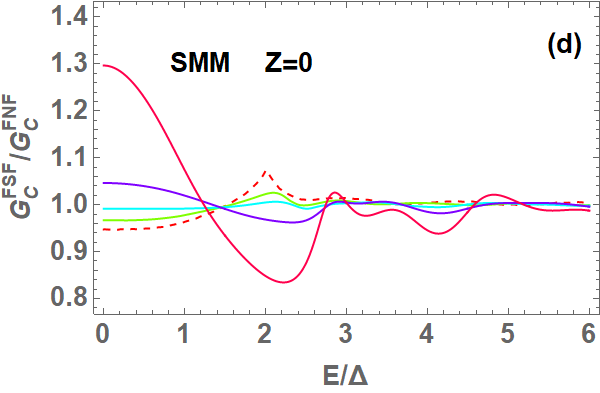}
\includegraphics[width=0.45\columnwidth, angle=0]{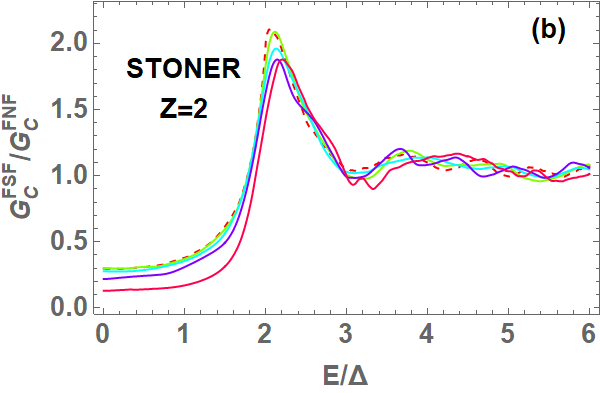}
\includegraphics[width=0.45\columnwidth, angle=0]{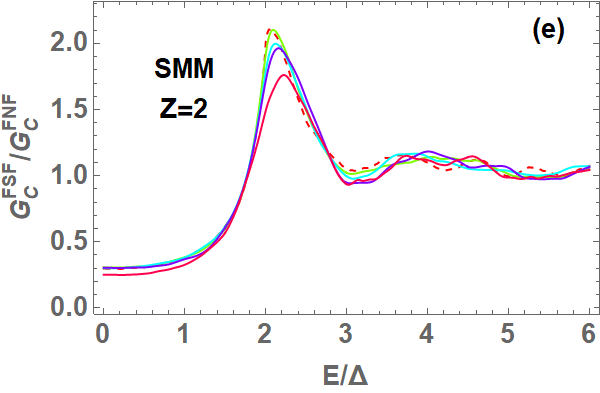}
\includegraphics[width=0.45\columnwidth, angle=0]{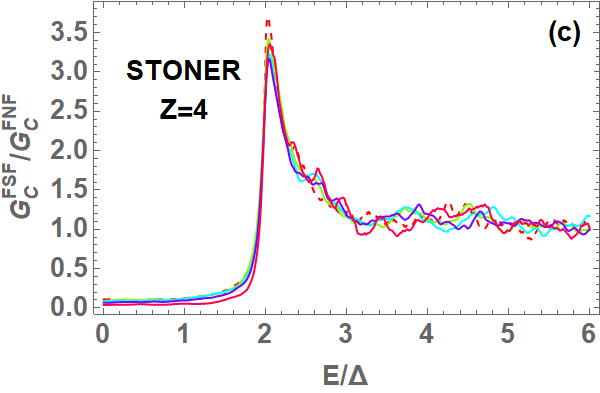}
\includegraphics[width=0.45\columnwidth, angle=0]{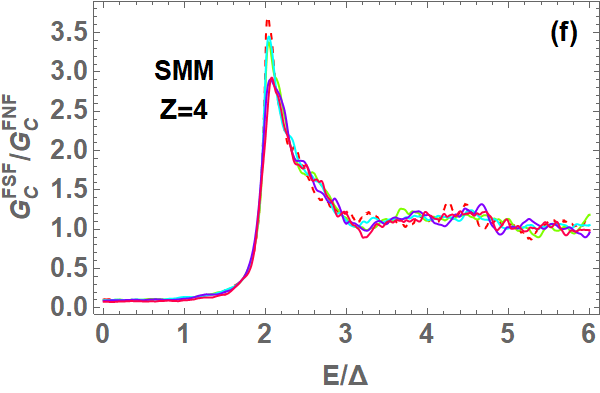}
\caption{Voltage bias dependence of the charge conductance integrated over all the allowed injection angles in the Stoner (a)-(c) and in the SMM case (d)-(f) at different values of barrier transparency: (a) and (d) for $\textsf{Z}=0$, (b) and (e) for $\textsf{Z}=2$,(c) and (f) for $\textsf{Z}=4$. The values of $M$ and the related color lines are the same as those used in Fig.~\ref{Fig_condC_th0}.}
\label{Fig_condC} 
\end{figure}

Similar results are obtained in the case of the charge conductance integrated over all possible injection angles, as shown by Fig.~\ref{Fig_condC}. Again we see a much larger low-bias weight in the SMM case than in the Stoner one for $\textsf{Z}=0$ and large magnetization, this difference tending to disappear at any $M$ when a lower and lower barrier transparency is considered. 

\subsubsection{Spin conductance}
\begin{figure}[htbp]
\includegraphics[width=0.45\columnwidth, angle=0]{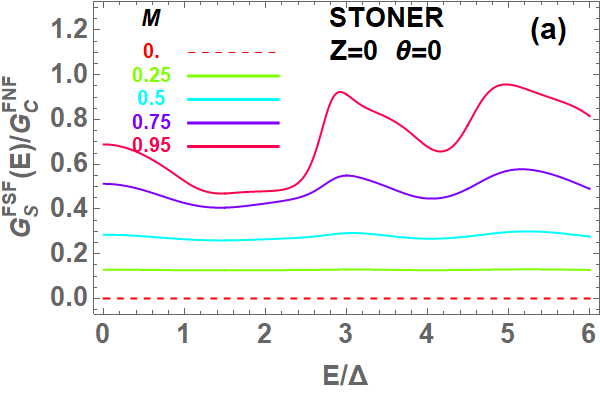}
\includegraphics[width=0.45\columnwidth, angle=0]{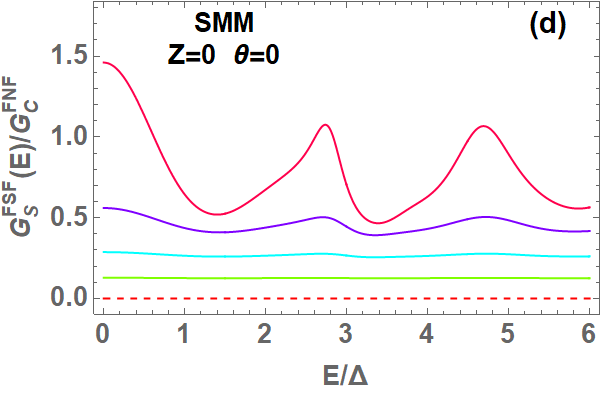}
\includegraphics[width=0.45\columnwidth, angle=0]{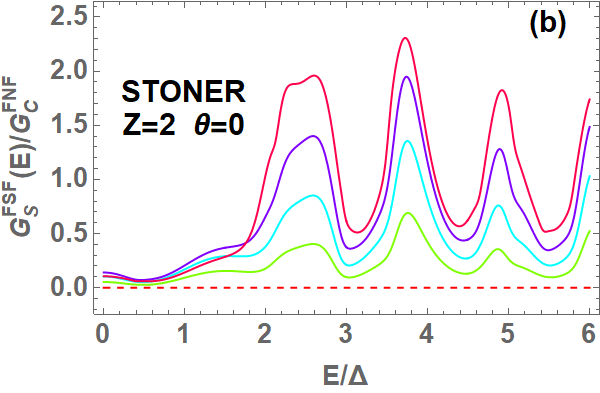}
\includegraphics[width=0.45\columnwidth, angle=0]{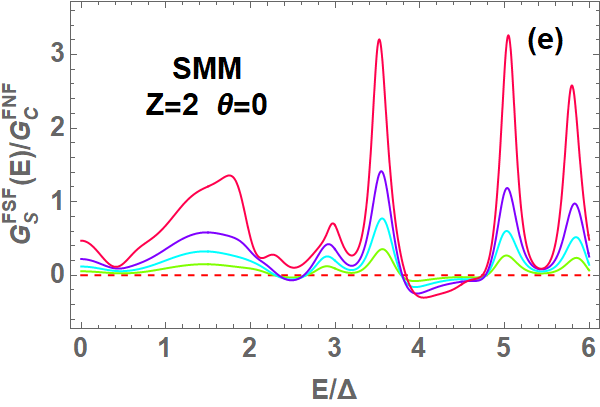}
\includegraphics[width=0.45\columnwidth, angle=0]{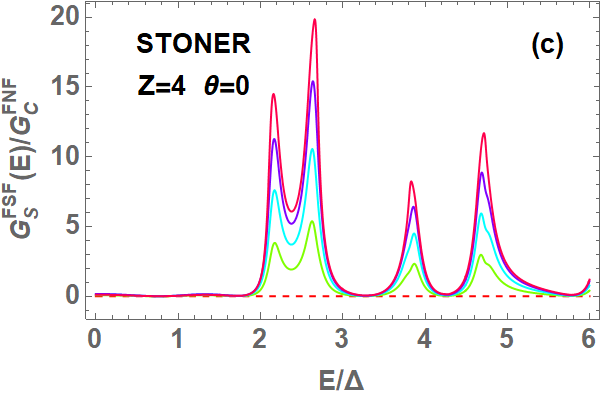}
\includegraphics[width=0.45\columnwidth, angle=0]{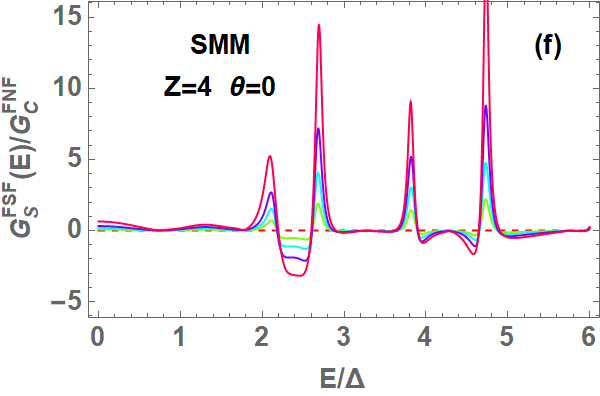}
\caption{Voltage bias dependence of the spin conductance at normal incident angle $(\theta=0)$ in the Stoner (a)-(c) and in the SMM case (d)-(f) at different values of barrier transparency: (a) and (d) for $\textsf{Z}=0$, (b) and (e) for $\textsf{Z}=2$,(c) and (f) for $\textsf{Z}=4$. The values of $M$ and the related color lines are the same as those used in Fig.~\ref{Fig_condC_th0}.}  
\label{Fig_condS_th0} 
\end{figure}

The behavior of the spin conductance is shown in Fig.~\ref{Fig_condS_th0} in the case of electron incidence normal to the interfaces. The results obtained for full transparency are shown in Fig.~\ref{Fig_condS_th0}\,(a) for the Stoner case and in Fig.~\ref{Fig_condS_th0}\,(d) for the SMM one. In particular, the spin conductance increases at all energies with increasing magnetization, this effect being at low bias much more pronounced in the SMM case than in the Stoner one. This can be explained noting that while with Stoner ferromagnets this trend comes from the asymmetrization of the probability coefficients and Fermi momenta of particles with opposite spin, in the SMM case the probability coefficients at $\theta=0$ and $\textsf{Z}=0$ are equal for spin-up and spin-down electrons, so that the amplitude of the spin conductance is positive and determined by the difference between the Fermi momenta of opposite spin electrons. As already pointed out, such a momentum difference becomes very significant in the SMM case for high magnetization values, due to the strong mass renormalization driving the ferromagnetic order. Furthermore, below the energy gap, the Andreev reflections are very strong and almost insensitive to polarization, thus allowing a magnitude of the spin conductance much larger in the SMM case than in the Stoner one,  in particular at low bias.  

In the presence of non transparent barriers and in particular in the tunnel limit, while the spin conductance of the F/S/F junction in the Stoner case systematically increases as a function of the ferromagnetic polarization [Figs.~\ref{Fig_condS_th0}\,(b) and \ref{Fig_condS_th0}(c)], in the SMM case there are ranges of the applied voltage bias where the spin conductance becomes negative, with an absolute value which slightly increases with the magnetization [Figs.~\ref{Fig_condS_th0}\,(e) and \ref{Fig_condS_th0}(f)]. 
Such feature comes from the asymmetry in the effective mass of opposite spin particles which leads to a larger velocity, and thus to a better transmission, of spin-down electrons compared to the spin-up ones [Figs.~\ref{CoeffSBZ2}(b) and \ref{CoeffSBZ2}(d)], thus reversing the role of majority spin electrons in the transmission process with respect to the Stoner case [Figs.~\ref{CoeffSTZ2}(b) and \ref{CoeffSTZ2}(d)]. It emerges especially at bias values above the superconducting gap and at large barrier values, since in such regime this effect does not compete anymore with the occurrence of Andreev reflections, and transport through the junction is fully dominated by the transmission of normal, non-superconducting quasiparticles.

\begin{figure}[htbp]
\includegraphics[width=0.43\columnwidth, angle=0]{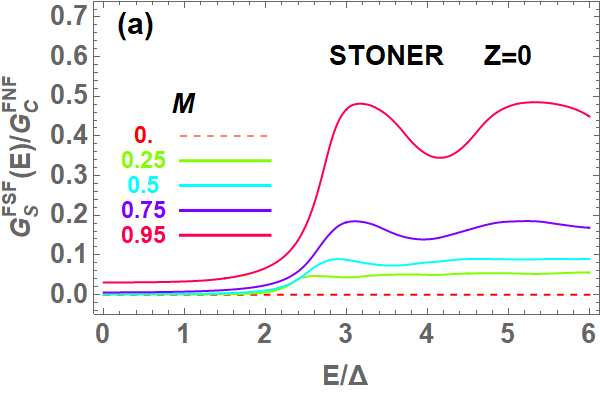}
\includegraphics[width=0.45\columnwidth, angle=0]{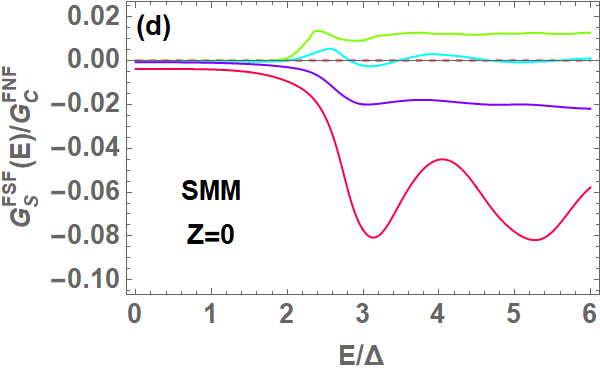}
\includegraphics[width=0.43\columnwidth, angle=0]{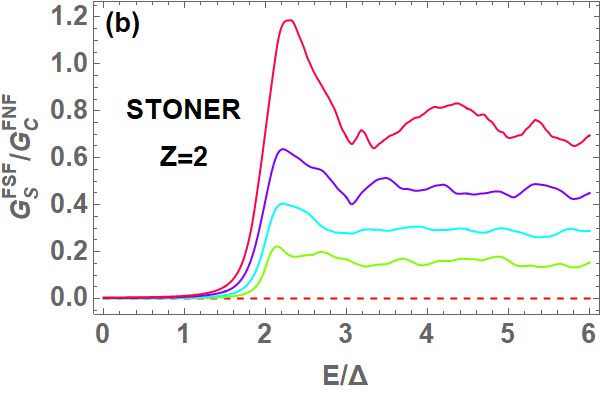}
\includegraphics[width=0.45\columnwidth, angle=0]{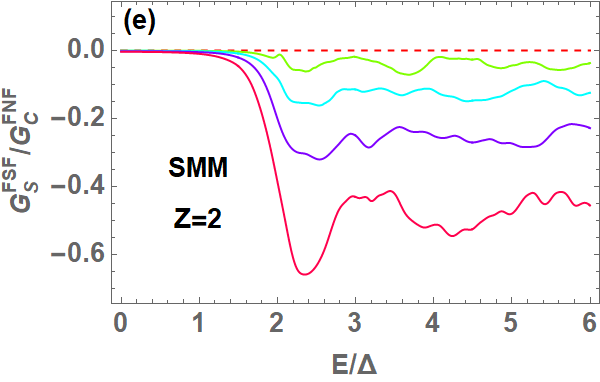}
\includegraphics[width=0.43\columnwidth, angle=0]{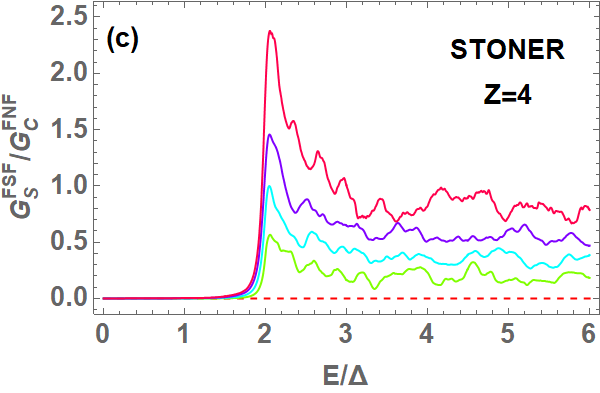}
\includegraphics[width=0.45\columnwidth, angle=0]{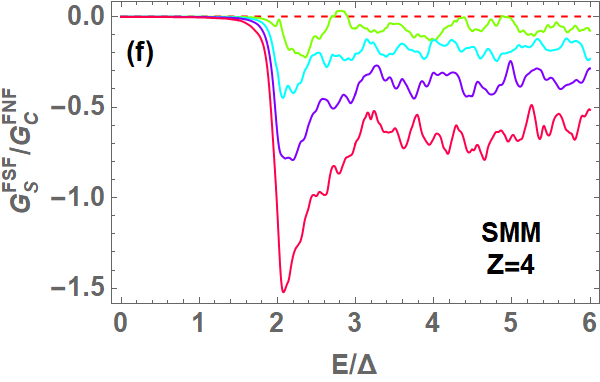}
\caption{Voltage bias dependence of the spin conductance integrated over all the allowed injection angles in the Stoner (a)-(c), and in the SMM case (d)-(f) at different values of barrier transparency: (a) and (d) for $\textsf{Z}=0$, (b) and (e) for $\textsf{Z}=2$,(c) and (f) for $\textsf{Z}=4$. The values of $M$ and the related color lines are the same as those used in Fig.~\ref{Fig_condC_th0}.}
\label{Fig_condS} 
\end{figure}

More pronounced differences between the Stoner and the SMM case are found when one considers the spin conductance integrated over all possible incidence directions [see Fig.~\ref{Fig_condS}].
In the Stoner case the spin conductance is always positive, for all values of the applied bias, the barrier transparency and the F layer magnetization. At low bias it almost vanishes, then becoming finite above the energy gap, with a magnitude which increases with the spin polarization [Figs.~\ref{Fig_condS}(a)- \ref{Fig_condS}(c)]. Below the gap, the spin conductance is negligible also in the case of SMM ferromagnetic layers, but differently from the Stoner case, at $\textsf{Z}=0$ and larger bias values it is positive at small $M$  [Fig.~\ref{Fig_condS}\,(d)], then becoming more and more negative as increasing values of $M$ are considered [Figs.~\ref{Fig_condS}\,(e) and \ref{Fig_condS}(f)]. For finite values of $\textsf{Z}$ we find a similar behavior, the only difference being that at large bias the spin conductance in the SMM case is always negative, regardless of the value of $M$.  

In order to understand this behavior, it is important to take into account that, according to the critical angle dependence on lead magnetization (see Appendix D),  while the injection cone of spin-up electrons is strongly suppressed by the magnetization, spin-down electrons can enter the junction at any injection angle. Consequently, in the integrated spin current, there is a strong competition between the contribution due to injected spin-up electrons, which is dominant at $\theta=0$ but restricted to a very limited angle range at increasing magnetization, and the contribution due to the injection of spin-down electrons, which is finite and sizable at all injection angles, in particular in the SMM case. Such competition gives also rise to very small values of the total spin conductance at bias lower than $2 \Delta$. 

We also notice that above $2\Delta$, the most significant contribution to the conductance comes from the excitation of normal quasiparticles, so that the sign of the spin conductance is dictated by the unbalance between the normal transmission of opposite spin particles. In the Stoner case one always gets positive values, increasing with the polarization, since at any magnetization, spin-up electrons have in the right ferromagnetic layer more accessible energy states than spin-down ones, due to the positive sign of the lead magnetization. On the contrary, the negative values found in the SMM case [see Figs.~\ref{Fig_condS}\,(d)-\ref{Fig_condS}(f)] come from the mass unbalance between spin-up and spin-down electrons, which favors the transmission of the faster down-spin electrons. On the other hand, in the bias regime below $2\Delta$, Andreev reflections counterbalance this tendency, since they have an approximately equal weight for spin-up and spin-down electrons, thus giving rise to a negligible spin current. 
Summarizing, while at biases below the gap the superconducting effects dominate via the Andreev scattering, above that value the dominant role is played by the ferromagnetic order which in the SMM case allows the spin-down current to dominate. 
 
Nevertheless, the presence of superconductivity induces an enhanced DOS at the gap edge, thus amplifying  the spin current for bias values of the order of $2 \Delta$:  around that value, the total charge and spin conductance are characterized by a peak, which is tunable through the polarization of the ferromagnetic leads, both in the Stoner and in the SMM case [see Figs.~\ref{Fig_condS}\,(c) and \ref{Fig_condS}(f)].

\begin{figure}[htbp]
\includegraphics[width=0.9\columnwidth, angle=0]{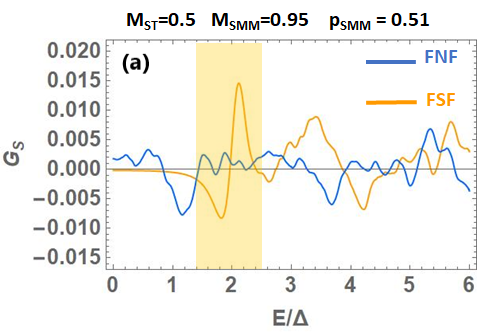}
\includegraphics[width=0.9\columnwidth, angle=0]{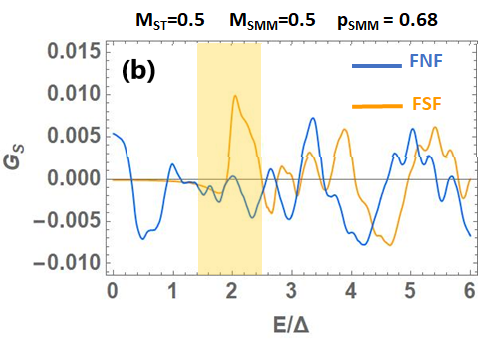}
\caption{Integrated spin conductance at $\textsf{Z}=2$ for two representative cases of F/N/F junctions, together with the corresponding F/S/F ones, where both ferromagnetic mechanisms occur with different polarizations (a) or different relative weights (b). Here $\textsf{p}_\textsf{SMM}$ is the weight of the SMM mechanism with respect to the Stoner one. The yellow-shaded region marks the energy window where distinctive features emerge in the F/S/F junction. }
\label{FigGsST@SMM} 
\end{figure}

The results we got show the emergence of peculiar effects within the context of superconductor-based magnetic double junctions. 
In this framework, many studies have been performed addressing spin and charge conductance in F/N/F structures, mainly in the context of spin-valve effects \cite{Johnson1985,Jedema2001,Jedema2002,Takahashi2003, Casanova2009}, as well as in F/S/F double junctions, where the effects of the relative orientation of the two ferromagnetic leads have been investigated. However, to our knowledge, no systematic study has been done in symmetric junctions on the spin response neither for F/N/F systems nor for F/S/F junctions as a function of the amplitude of the magnetization of the ferromagnetic leads, exploring the role of different metallic ferromagnets. In particular, we are not aware of measurements demonstrating a sign reversal for the spin conductance as a function of the ferromagnet polarization for normal incidence, or negative values of the integrated spin conductance, in symmetric F/N/F junctions. We argue that this may be ascribed to the fact that in realistic materials the mechanism responsible for ferromagnetism could be a combination of both magnetic exchange (Stoner) and kinetic (mass mismatch) modifications of the spin dependent electronic structure. Since the two mechanisms give rise to opposite spin currents, their simultaneous presence in a given material can be responsible in the corresponding F/N/F junction for a spin conductance that, on the average, is vanishing  or difficult to detect. In 
Fig.\ref{FigGsST@SMM}, we show the results of the integrated spin conductance at $\textsf{Z}=2$ for two representative cases of F/N/F junctions, together with the corresponding F/S/F ones, where both ferromagnetic mechanisms are active with different polarizations [Fig.\ref{FigGsST@SMM}(a)] or different relative weights [Fig.\ref{FigGsST@SMM}(b)]. We have denoted by $\textsf{p}_\textsf{SMM}$ the weight of the spin mismatch mechanism with respect to the Stoner one, and calculated the total spin conductance as
\[
G_S=(1-\textsf{p}_\textsf{SMM})\,G_S^{Stoner}+\textsf{p}_\textsf{SMM}\,G_S^{\textsf{SMM}}\; .
\]
We observe that in the F/N/F case, the spin conductance is substantially featureless with incoherent oscillations in energy. On the other hand, the use of the superconductor allows to exploit its characteristic energy scale, associated with the superconducting gap $\Delta$, and focus on a specific energy window in the analysis of the transport properties. In our case the latter corresponds to the yellow-shaded region in Fig.\ref{FigGsST@SMM}, where the energy dependence of the spin conductance is marked by distinctive features around $\textsf{E}=2 \Delta$. 
In particular, we find that when the SMM mechanism is predominant, the spin conductance exhibits a dip-peak structure developing from negative to positive values that tends to evolve towards a positive single peak as the weight of the Stoner mechanisms gradually increases. Based on that, we expect that a measurement of the spin conductance of a F/S/F junction may provide relevant hints on the extent to which the two mechanisms compete between each other.

\section{Conclusions}
We have presented a study of the transport phenomena in a clean F/S/F junction with parallel magnetization in the two F layers, making a comparison between the case of Stoner-type ferromagnetic layers with the one where ferromagnetism is driven by an asymmetric mass renormalization of carriers with opposite spin. 
We have shown that charge and spin transport in this junction exhibits different features depending on the mechanism which is responsible for the ferromagnetism, in the case of perpendicular injection as well as when considering the integrated behavior over all the allowed injection directions.

In particular, for transparent barriers, Andreev reflections are more robust in the SMM case than in the Stoner one as the magnetization in the F layers is increased. As a consequence, with SMM ferromagnets transport through the junction is characterized by a significant amplification of the charge conductance with respect to the F/N/F case, which is not observed in the Stoner case. Then, the spin conductance of the SMM junction monotonously increases with the ferromagnetic exchange for particles injected perpendicularly to the barriers, in opposition to the nonmonotonous behavior versus magnetization found in the Stoner case. Finally, in the tunnel limit, while the charge conductance assumes values which weakly depend on the ferromagnetic mechanism, the spin conductance exhibits opposite signs in the SMM and in the Stoner case at large applied bias. In both cases, the superconducting pairing enhances the amplitude of the spin current close to the gap edge.

So far, several magnetic materials have been found to exhibit properties that cannot be framed exclusively within a Stoner scenario \cite{Okimoto, hexabor, higashiguchi2005}. This often happens in the cases of half-metal ferromagnets, where the almost full degree of spin polarization develops in regimes where a mass mismatch is clearly distinguishable and high polarization values cannot be explained in terms of the Stoner mechanism only \cite{sun2013}. For these systems, a theoretical analysis where the role played by spin-dependent electron masses is explicitly taken into account is likely to be required. Within this context, the different behavior predicted in the Stoner and in the SMM cases may provide useful indications on the nature of the mechanism originating the ferromagnetic order in a given ferromagnetic material. Given the usual limitations in the experimental realization of heterostructures, the possibility of selecting the magnetization mechanism, in this way controlling the spin of the carrier responsible for transport, may turn out to be useful in the design of electronics and spintronics devices.  
In this context, it may also be interesting to investigate to what extent the use of SMM ferromagnets, instead of Stoner-like ones, affects the behavior of a F/S/F junction when treated as a spin valve.

We finally point out that in the present analysis the three subsystems are all considered in the clean limit. Concerning the dependence of the results on the sample purity, we expect that in the regime of weak disorder and in the absence of spin-flip scattering in the superconductor, disorder is mainly affecting the coherence length, which becomes $\xi_D \sim \sqrt{\hbar D/\Delta}$ with $D$ being the diffusion constant and $\Delta$ the superconducting gap. This renormalization implies that, with respect to the investigated clean configuration, similar results are expected by scaling the thickness $L$ of the superconductor. 
However, when the disorder introduces strong energy relaxation processes and spin flip scattering, our results are no longer valid and a different approach is required to deal with the spin diffusion in the superconductor. For instance, the analysis performed in Ref.\onlinecite{Morten2004}, involving Stoner type ferromagnets and spin flip scattering in the superconductor, showed that the spin current is suppressed at bias below the superconducting energy gap, and a massive spin flip occurs at energies close to the gap. These processes can of course modify the obtained results near the gap edge. A complementary approach including an inelastic transport regime has been proposed in Refs.\onlinecite{Yamashita2002, Takahashi2003}, where it is shown that in a F/S/F structure the superconductor becomes a low-carrier system for spin transport, due to the opening of the gap, and thus the accumulation spin signal is greatly enhanced with respect to a non-superconducting layer. On the basis of this result, we argue that inelastic processes can help to distinguish the spin signal when going from the normal to the superconducting phase, and that might be applicable for both Stoner and SMM ferromagnets.  
Along this line, we point out that to the best of our knowledge, all the performed studies in the presence of disorder deal with Stoner ferromagnetic leads, while an investigation of the effects of disorder in the case of SMM-based ferromagnets is still lacking. This problem goes beyond the scope of this work and will be considered in future investigations.

\section*{Appendix A}
In this Appendix we complement Eqs.~(\ref{psi1E})-(\ref{psi2E}) reporting the expression of the wave functions in the three regions of the junction for injections other than the one of electrons with spin $\sigma$ from the left F side.

For the injection of a hole with energy $\varepsilon$ and spin $\sigma$ from the left F side, the wave functions in the three regions of the junction are: 

\begin{eqnarray}
\nonumber \psi_{h\sigma \textsf{L}}^{\textsc{F}}(z)&=&
\left(
\begin{array}{l}
 0\\
 1 
\end{array}
\right) e^{-i q_{\sigma} z \ cos\theta_{\sigma}}
+ \textsf{a}_{h \bar{\sigma}}
\left(
\begin{array}{l}
1\\
0
\end{array}
\right)
e^{-i q_{\bar{\sigma}} \ z \ cos\theta^{A}_{\bar{\sigma}}} \\
& & + \, \textsf{b}_{h \sigma}
\left(
\begin{array}{l}
0\\
1
\end{array}
\right)
e^{i q_{\sigma} z \ cos\theta_{\sigma}} 
\label{psi1H}
\end{eqnarray}
for $z<0$;
\begin{eqnarray}
\nonumber 
\psi^{\textsc{S}}_{h \sigma} (z)&=& 
\alpha_{h\sigma}
\left(
\begin{array}{l}
 u_0\\
 v_0 
\end{array}
\right) 
e^{i k_e z \ cos\theta_{\sigma e}^{\textsc{S}}}
\\ \nonumber & & 
+ \, \beta_{h \bar{\sigma}}
\left(
\begin{array}{l}
v_0\\
u_0 
\end{array}
\right)
e^{- i k_h z \ cos\theta_{\sigma h}^{\textsc{S}}}
\\ 
\nonumber & & + \,
\gamma_{h\sigma}
\left(
\begin{array}{l}
 u_0\\
 v_0 
\end{array}
\right) 
e^{-i k_e z \ cos\theta_{\sigma e}^{\textsc{S}}}
\\ 
& & + \, \eta_{h \bar{\sigma}}
\left(
\begin{array}{l}
v_0\\
u_0 
\end{array}
\right)
e^{i k_h z \  cos\theta_{\sigma h}^{\textsc{S}}}
\label{psiSH}
\end{eqnarray}
for $0<z<L$; 
\begin{eqnarray}
\nonumber \psi_{h\sigma \textsf{R}}^{\textsc{F}}(z)&=&
\textsf{c}_{h \bar{\sigma}}
\left(
\begin{array}{l}
0\\
1
\end{array}
\right)
e^{-i q_{\bar{\sigma}}  z \ cos\theta^{T}_{\sigma} }
\qquad\qquad \\ && \qquad\quad + \; \textsf{d}_{h \sigma}
\left(
\begin{array}{l}
1\\
0 
\end{array}
\right)
e^{ i q_{\bar{\sigma}} z \ cos\theta^{T}_{\bar{\sigma}} }
\label{psi2H}
\end{eqnarray}
for $z>L$.


The expressions of the wave functions corresponding to the injection of an electron with energy $\varepsilon$ and spin $\sigma$ from the right F side can be obtained from Eqs.~(\ref{psi1E})-(\ref{psi2E}) by reversing the sign of all wavevectors.
The same substitution can be applied to Eqs.~(\ref{psi1H})-(\ref{psi2H}) to get the wave functions corresponding to the injection of a hole with energy $\varepsilon$ and spin $\sigma$ from the right ferromagnet. Due to the symmetry of the problem with respect to the superconducting layer, the probability amplitudes of the scattering processes corresponding to particle injection from the right side are equal to those of the corresponding processes for the same particle injection from the left side.

\section*{Appendix B}
In this Appendix we derive the probability current conservation through the junction, showing where the mass asymmetry condition enters, and how it affects the expression of the probabilities associated with the scattering processes occurring in the junction.

Denoting by $P_{\sigma}(\textbf{r},t)=|\Psi_{\sigma}(\textbf{r},t)|^{2}$ the probability density to find a particle at a given time $t$ in the volume element $d\textbf{r}$ around the position $\textbf{r}$, we have
\begin{equation}
\frac{d}{dt} \int d\textbf{r} \, P_{\sigma}(\textbf{r},t)=0 \; ,
\label{cons}
\end{equation}
with the integrals extended to the whole space. Using the Schr\"{o}dinger equations for the spinors $\Psi_{\sigma}(\textbf{r},t)$ and $\Psi_{\sigma}^{*}(\textbf{r},t)$  
\begin{eqnarray}
&&\imath\hbar\frac{\partial}{\partial t}\Psi_{\sigma}(\textbf{r},t)=H_{\sigma}^{BdG}\Psi_{\sigma}(\textbf{r},t)\\
-&&\imath\hbar\frac{\partial}{\partial t}\Psi_{\sigma}^{*}(\textbf{r},t)=H^{BdG}_{\sigma}\Psi_{\sigma}^{*}(\textbf{r},t)
\end{eqnarray}\\
and taking into account that, due to the independence of the Hamiltonian of the time coordinate, the time dependence of the wave function can be factorized, we finally get:
\begin{eqnarray}
\frac{\partial P_{\sigma}(\textbf{r},t)}{\partial t} & = & \frac{2}{\hbar}\,{\rm Im}\{u_{\sigma}^{*}(\textbf{r})\hat{H}_{\sigma} (\textbf{r})u_{\sigma}(\textbf{r})-v^{*}_{\bar{\sigma}}(\textbf{r})\hat{H}^{*}_{\bar{\sigma}}(\textbf{r})v_{\bar{\sigma}}(\textbf{r})\} \nonumber \\
& = &\hbar \, {\rm Im}(u^{*}_{\sigma}(\textbf{r})\frac{\hat{\textbf{p}}^{2}}{m_{\sigma}}u_{\sigma}(\textbf{r})-v_{\bar{\sigma}}^{*}(\textbf{r})\frac{\hat{\textbf{p}}^{2}}{m_{\bar{\sigma}}}v_{\bar{\sigma}}(\textbf{r})) \; .
\end{eqnarray}
In the absence of magnetic field, the momentum operator is $\hat{\textbf{p}}= -\imath \hbar {\boldsymbol\nabla}$ so that one finally gets the continuity equation
\begin{equation}
\frac{\partial}{\partial t}P_\sigma(\textbf{r},t)+{\boldsymbol\nabla}\cdot \textbf{J}_{\sigma}(\textbf{r})=0
\end{equation}
with the probability density current $\textbf{J}_{\sigma}(\textbf{r})$ given by
\begin{eqnarray}
\textbf{J}_{\sigma}(\textbf{r}) & = & {\rm Im} \left[\frac{\hbar}{m_{\sigma}} u_{\sigma}^{*}(\textbf{r}){\boldsymbol\nabla}u_{\sigma}(\textbf{r})-\frac{\hbar}{m_{\bar{\sigma}}} v_{\bar{\sigma}}^{*}(\textbf{r}){\boldsymbol\nabla}v_{\bar{\sigma}}(\textbf{r})\right] \; . \nonumber \\
& &
\label{jp}
\end{eqnarray}

The expression of $\textbf{J}_{\sigma}$ clearly shows that an asymmetry in the effective mass of electrons with opposite spin also enters the formal expression of the probability density current. 
Such expression can be used to derive the probability coefficients associated with the scattering processes taking place in the junction. 
In particular, from the probability density current conservation it follows that the total current flowing through a surface enclosing the whole system is zero.  
By considering the probability density current flowing through the interfaces, and thus along the $z$-direction, there are the following contributions: the probability density current $J_{\sigma}^{I}$ for the incident particles with spin $\sigma$, and the associated reflected and transmitted currents $J_{\sigma}^{R}$ and $J_{\sigma}^{T}$, respectively. They are related by the following equation: 
\begin{equation}
 J_{\sigma}^{I}+J_{\sigma}^{R}= J_{\sigma}^{T} \; .
\label{consJ}
\end{equation}
By using the wave functions defined in each region of the junction in the representative case of injected electrons with spin $\sigma$, we have that the projections of the currents in the direction perpendicular to the interfaces are:
\begin{eqnarray}
J_{\sigma}^{I}&=&\frac{\hbar}{m_{\sigma}}q_{\sigma} \cos\theta_{\sigma}\\
J_{\sigma}^{R}&=&-\frac{\hbar}{m_{\sigma}}|\textsf{b}_{e\sigma}|^2 q_{\sigma} \cos\theta_{\sigma} -\frac{\hbar}{m_{\bar{\sigma}}}|\textsf{a}_{e\bar{\sigma}}|^2 q_{\bar{\sigma}} \cos\theta^A_{\bar{\sigma}}\\
 J_{\sigma}^{T} &=&
  \frac{\hbar}{m_{\sigma}}
|\textsf{c}_{e\sigma}|^2 
q_{\sigma} \cos\theta^T_{\sigma} + \frac{\hbar}{m_{\bar{\sigma}}}
|\textsf{d}_{e\bar{\sigma}}|^2 q_{\bar{\sigma}} \cos\theta^T_{\bar{\sigma}}
 \; .
\end{eqnarray}
By applying Eq.(\ref{consJ}) and dividing all the terms by the injected current, we get the relation 
\begin{equation}
1=\textsf{A}_{\sigma}+\textsf{B}_{\sigma}+\textsf{C}_{\sigma}+\textsf{D}_{\sigma}
\end{equation}
which allows to define the probability coefficients for generic injected particles $\textsf{p}$:  
$\textsf{A}_{\textsf{p}\sigma}$, $\textsf{B}_{\textsf{p}\sigma}$, $\textsf{C}_{\textsf{p}\sigma}$, $\textsf{D}_{\textsf{p}\sigma}$
(being $\textsf{p}=e,h$ for electrons and holes, respectively) as reported in Eqs.~(\ref{eqApr})-(\ref{eqDpr}).

\section*{Appendix C}

In this Appendix we present the detailed derivation of the junction conductance, following an extension of the original BTK approach~\cite{BTK} to the case of a F/S/F double junction with Stoner ferromagnets \cite{yamashita03}. 

As for the case of the single junction, we calculate the conductance in the left ferromagnetic side, where the current flow does not include supercurrents and the calculation is therefore more convenient. Taking into account the results presented in Appendix B, the charge current flowing in the presence of an applied bias $V$ from the left to the right side of the junction can be calculated as: 
\begin{eqnarray}
J_\sigma & = & \hbar \  {\rm Im}
\sum_{l, \sigma, \epsilon } \textsc{q}_l \left[ \frac{f_{\epsilon}}{m_{l \sigma}} \, u_{l \sigma}^{*} 
\frac{\partial}{\partial z} u_{l \sigma}
 + \right.  
 \nonumber \\  && \qquad\qquad + \left.
 \frac{(1-f_{\epsilon})}{m_{l \bar{\sigma}}} \, v_{l \sigma}^{*} \frac{\partial}{\partial z} v_{l \sigma} \right] \; .
\end{eqnarray} 
Here $f_{\epsilon}$ is the Fermi distribution function and the subscript $l=1,4$ refers to the four possible injection processes described in the main text and graphically shown in Fig.~\ref{figprocessiok}. We thus have $\textsc{q}_{1}=\textsc{q}_{3}=e$ and $\textsc{q}_{2}=\textsc{q}_{4}=-e$, $e \, (<0)$ being the electron charge.

When a bias potential $V$ is applied between the two F leads, by taking into account that the summation on the energies involves the energy levels of the side where particles are injected, we can write the following expressions for the normal components of the spin-dependent currents in the left F side:

\begin{eqnarray}
\nonumber J_\sigma & = & e \hbar \sum_{\theta,\epsilon \in F_1} \left[ \frac{q_{\sigma}}{m_{\sigma}}\cos\theta(1-|\textsf{b}_{e \sigma}|^2)f\left(\epsilon - \frac {e V}{2}\right) \right. \\\nonumber && - \left. \frac{q_{\bar{\sigma}}}{m_{\bar{\sigma}}}\cos\theta^{A}_{\bar{\sigma}}|\textsf{a}_{e \bar{\sigma}}|^2\left[1- f\left(-\epsilon -\frac{e V}{2}\right)\right]\right\}
\\ \nonumber &&- \, e \hbar \sum_{\theta,\epsilon \in F_1} \left\{-\frac{q_{\bar{\sigma}}}{m_{\bar{\sigma}}}\cos\theta^\textsf{A}_{\bar{\sigma}}\,|\textsf{a}_{h \bar{\sigma}}|^2 f\left(\epsilon -\frac{e V}{2}\right) \right. 
\\ \nonumber &&+ \left.  \frac{q_{\sigma }}{m_{\sigma}} \cos\theta\ (1-|\textsf{b}_{h \sigma}|^2)\left[1- f\left(-\epsilon -\frac{e V}{2}\right)\right]\right\}
 \\ \nonumber  &&+ \, e \hbar \sum_{\theta,\epsilon \in F_2} \left\{-\frac{q_{\sigma}}{m_{\sigma}}\cos\theta^T_{\sigma}\,|\tilde{\textsf{c}}_{e \sigma}|^2 f\left(\epsilon +\frac{e V}{2}\right)  \right. 
 \\  \nonumber && -\left.  \frac{q_{\bar{\sigma}}}{m_{\bar{\sigma}}}\cos\theta^T_{\bar{\sigma}}|\tilde{\textsf{d}}_{e \bar{\sigma}}|^2\left[1- f\left(-\epsilon + \frac{e V}{2}\right)\right]\right\}
 \\ \nonumber && - \, e \hbar \sum_{\theta,\epsilon \in F_2} \left\{\frac{q_{\sigma}}{m_{\sigma}}\cos\theta^T_{\sigma} \,|\tilde{\textsf{c}}_{h \sigma}|^2 \left[1- f\left(-\epsilon- \frac{e V}{2}\right)\right] \right. \\ 
&& + \left.  \frac{q_{\bar{\sigma}}}{m_{\bar{\sigma}}}\cos\theta^T_{\bar{\sigma}}|\tilde{\textsf{d}}_{h \bar{\sigma}}|^2 f\left(\epsilon-\frac{e V}{2}\right)\right\} \; .
\end{eqnarray}
Taking into account that $1-f(\epsilon)=f(-\epsilon)$ and assuming that the two ferromagnets are identical, so that the probability amplitudes of scattering processes due to particles injected from the right side are equal to those for the same particle injection from the left side (in particular $\tilde{\textsf{c}}_{\textsf{p}\sigma}=  \textsf{c}_{\textsf{p}\sigma}$, and $\tilde{\textsf{d}}_{\textsf{p}\bar{\sigma}}=  \textsf{d}_{\textsf{p}\bar{\sigma}}$), it is possible to write the current as 

\begin{equation}
\nonumber J_\sigma = J_{e\sigma} +J_{h\sigma }
\end{equation}
with
\begin{eqnarray}
\nonumber J_{e\sigma}&=& e \hbar \sum_{\theta,\epsilon \in F_1}
 \frac{q_{\sigma}}{m_{\sigma}}\cos\theta
 \left[
 f\left(\epsilon - \frac {e V}{2}\right)
 \left(
 1-|\textsf{b}_{e \sigma}|^2 \right. \right. \\
 &-& \left. \left. \frac{q_{\bar{\sigma}}}{q_{\sigma}}
 \frac{m_{\sigma}} {m_{\bar{\sigma}}} \frac{\cos\theta_{\bar{\sigma}}^{T}}{\cos\theta} |\textsf{d}_{e \bar{\sigma}}|^2 
 \right) -  f\left(\epsilon + \frac {e V}{2}\right) \right.
 \\ \nonumber && \left.
 \left(
  |\textsf{a}_{e \bar{\sigma}}|^2  \frac{q_{\bar{\sigma}}}{q_{\sigma}}
 \frac{m_{\sigma}} {m_{\bar{\sigma}}} \frac{\cos\theta_{\bar{\sigma}}^{A}}{\cos\theta} +  |\textsf{c}_{e \sigma}|^2 \right) 
 \right]  
 \\ \nonumber &=&   e \hbar \sum_{\theta,\epsilon \in F_1}  \frac{q_{\sigma}}{m_{\sigma}}\cos\theta
 \left[
 f(\epsilon - \frac {e V}{2})
 \left(
 1- \textsf{B}_{e \sigma} -  \textsf{D}_{e \sigma} \right) \right.
 \\ & & - \left.  f\left(\epsilon + \frac {e V}{2}\right)
 \left(
\textsf{A}_{e \sigma} +  \textsf{C}_{e \sigma} \right) 
 \right]  
 \end{eqnarray}
and
\begin{eqnarray}
\nonumber J_{h\sigma}&=&- e \hbar \sum_{\theta,\epsilon \in F_1}
 \frac{q_{\sigma}}{m_{\sigma}}\cos\theta
 \left[
 f\left(\epsilon + \frac {e V}{2}\right)
 \left(
 1-|\textsf{b}_{h \sigma}|^2 - 
\right. \right. \\  && \nonumber
 \left. \left.  
   \frac{q_{\bar{\sigma}}}{q_{\sigma}}
  \frac{m_{\sigma}} {m_{\bar{\sigma}}} 
  \frac{\cos\theta_{\bar{\sigma}}^{T}}{\cos\theta} |\tilde{\textsf{d}}_{h \sigma}|^2 
  \right) \right.  \\ \nonumber && \left. -  f\left(\epsilon - \frac {e V}{2}\right)   \left(
  |\textsf{a}_{h \bar{\sigma}}|^2  \frac{q_{\bar{\sigma}}}{q_{\sigma}}
\frac{m_{\sigma}} {m_{\bar{\sigma}}} \frac{\cos\theta_{\bar{\sigma}}^{A}}{\cos\theta} +  |\tilde{\textsf{c}}_{h \sigma}|^2 \right) 
 \right]   
 \\ \nonumber & = & - e \hbar \sum_{\theta,\epsilon \in F_1}  \frac{q_{\sigma}}{m_{\sigma}}\cos\theta
 \left[
 f\left(\epsilon + \frac {e V}{2}\right)
 \left(
 1- \textsf{B}_{h \sigma} - \textsf{D}_{h \sigma} \right) \right. 
 \\
 && \left.  -  f\left(\epsilon - \frac {e V}{2}\right)
 \left(
\textsf{A}_{h \sigma} + \textsf{C}_{h \sigma} \right) 
 \right]  
 \end{eqnarray}
From the conservation of the probability current, we finally get:
\begin{eqnarray}
\nonumber J_\sigma & = & e  \hbar \sum_{\theta,\epsilon \in F_1}  \frac{q_{\sigma}}{m_{\sigma}}\cos\theta \left[ f(\epsilon - \frac {e V}{2}) - f(\epsilon + \frac {e V}{2}) \right] \\
&& \times \left( \textsf{A}_{e \sigma}  +  \textsf{C}_{e \sigma} + \textsf{A}_{h \sigma} +  \textsf{C}_{h\sigma} \right) \, .
\end{eqnarray}
By considering that the sum over the allowed energies and injection angles can be written as 
\[
\sum_{\theta,\epsilon \,\in F_1}=  \int_{0}^{\pi/2} d\theta \int d\epsilon \, N^{\textsc{F}}_{\sigma}(\epsilon)
\]
and that the spin dependent density of states $N^{F}_{\sigma}(\epsilon)$ in the two equivalent ferromagnets in the 2D case is constant and equal to $N^{F}_{\sigma}(\epsilon)=m_{\sigma}/(2 \pi \hbar^{2})$, the spin-dependent charge conductance can be written as:
\begin{eqnarray}
G_\sigma(\textsf{E})= \frac{dJ_\sigma}{dV} = \int_{0}^{\pi/2} d\theta  \ \textsc{G}_{\sigma}(\theta,\textsf{E})
\end{eqnarray}
where $\textsf{E}=e V$ and
\begin{eqnarray*}
 \textsc{G}_{\sigma}(\theta,\textsf{E}) = G_0 \ \tilde{q}_{\sigma} \cos\theta \left(\textsf{A}_{e \sigma}+ \textsf{C}_{e \sigma} + \textsf{A}_{h \sigma}+ \textsf{C}_{h\sigma}\right)\left|_{\frac{\textsf{E}}{2},\theta}\right. \; .
\end{eqnarray*}
Here $\tilde{q}_{\sigma}= q_{\sigma}/ q_{F}$ and $G_0=\frac{e^2  q_F }{\pi \hbar} $ is the conductance of the junction when the three layers are all in the normal state.

\section*{Appendix D}

It is known that the measured conductance takes contributions from a range of incidence angles which depends on the conditions under which experiments are performed.  
For electrons and holes injected from the left side, two limiting angles have to be considered:
(a) the incident angle above which local Andreev reflections can not occur, and (b) the limiting angle of incidence for the transmission into the superconductor. 
Such limiting angles can be derived from the application of the conservation law given by Eq.~(\ref{BCangle2}). 

\begin{figure}[h!]
\centering
\includegraphics[width=0.98\columnwidth, angle=0]{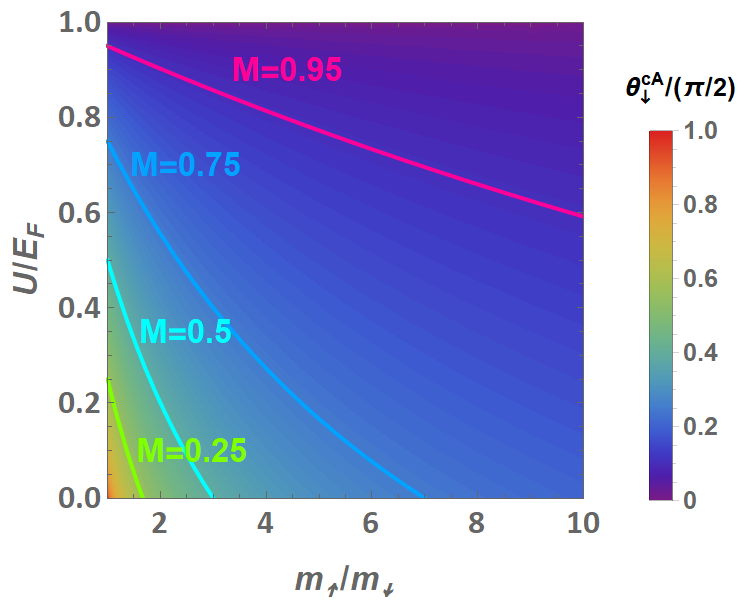}   
\caption{Density plot of the critical angle for the local Andreev reflections of spin-up electrons (and holes) injected from the left side (or equivalently from the right side since we have considered the two ferromagnetic leads fully identical), as a function of the microscopic parameters $X=U/E_F$ and $Y=m_{\uparrow}/m_{\downarrow}$, which are assumed to be the same in the two ferromagnetic leads. 
Isomagnetization lines for different valued of the magnetization are reported. Each of these lines also corresponds to a fixed critical angle value. No limitation to the injection direction of spin-down electrons (holes) holds when the magnetization is assumed positive.}
\label{figthetaCRA}
\end{figure}

\begin{figure}[h!]
  \centering
 \includegraphics[width=0.98\columnwidth, angle=0]{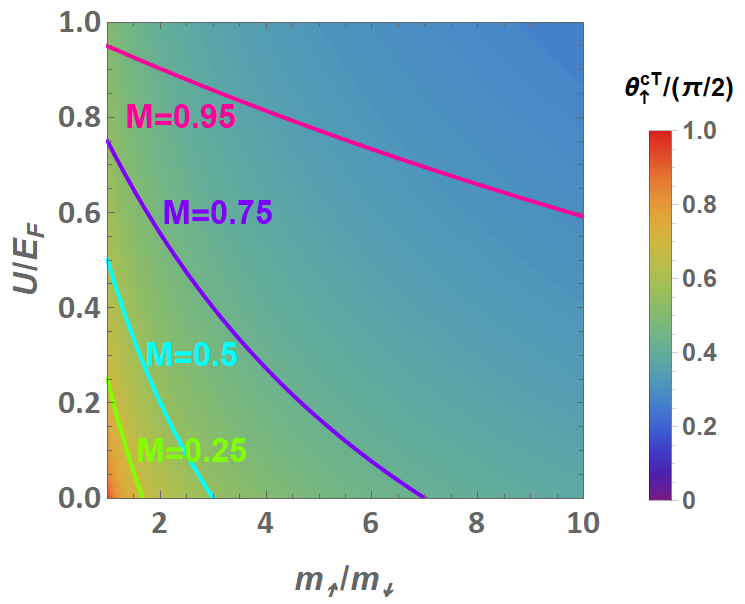}
 \caption{Density plot of the critical angle for trasmission to $S$ of spin-up electrons (and holes) injected from the left side (or equivalently from the right side, since we have considered the two ferromagnetic leads  identical), as a function of the microscopic parameters $X=U/E_F$ and $Y=m_{\uparrow}/m_{\downarrow}$. Isomagnetization lines for different valus of the magnetization are also reported. No limitation to the injection direction of down-spin electrons (holes) holds when the magnetization is assumed positive.}
\label{figthetaCRT}
\end{figure}

 \begin{figure}[h!]
\centering
 \includegraphics[width=0.98\columnwidth, angle=0]{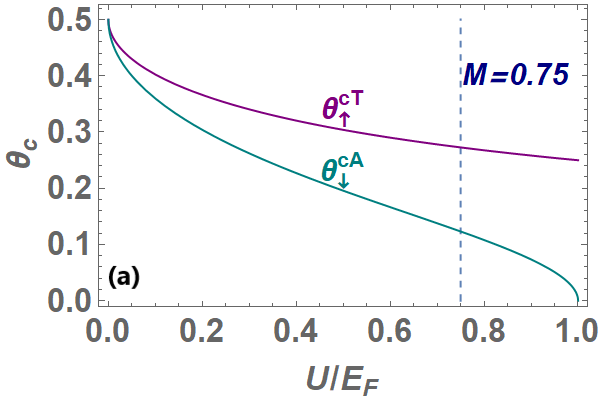}  
 \includegraphics[width=0.98\columnwidth, angle=0]{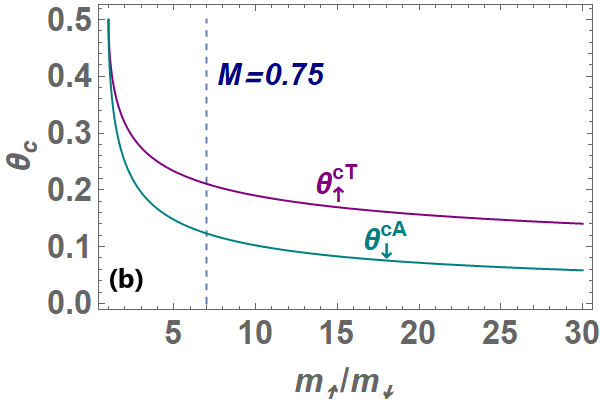}
\caption{Critical injection angles for the Andreev reflections of spin-up electrons (and holes) entering from the left side (or equivalently from the right side, since we have considered identical ferromagnetic leads), as functions of the microscopic parameters $X=U/E_F$ and $Y=m_{\uparrow}/m_{\downarrow}$, in the Stoner (a) and in the SMM case (b). The dotted line in the two panels indicates the magnetization value $M=0.75$.}
\label{figthetaCRAvsCRT}
\end{figure}

In the case where the two ferromagnets are identical, and assuming that the magnetization amplitude is positive, we find that in the case of spin-down injected particles, both local Andreev reflections and particle transmissions into the S layer can occur independently of the injection direction. Indeed, according to Eq.~(\ref{BCangle2}), the scattering angles defining the direction of the Andreev reflections and the transmission in S are

\begin{eqnarray}
\theta_{\uparrow}^{A} & = & \arcsin\left[\frac{q_{\downarrow}}{q_{\uparrow}} \sin \theta_{\downarrow} \right]\label{thetaA_DW}\\
\theta_{\downarrow}^{T} & = & \arcsin\left[\frac{q_{\downarrow}}{k_F^S} \sin \theta_{\downarrow} \right]
\label{thetaT_DW}
\end{eqnarray}
where $\theta_{\sigma}$ is the angle of the particles which are injected with  spin ${\sigma}$. When the lead magnetization is assumed positive, $q_{\downarrow}$ is less than $q_{\uparrow}$, and is also less than $k_F^S$. Consequently, in the right side of Eqs.~(\ref{thetaA_DW})-(\ref{thetaT_DW}), the argument of the arcsine functions is always less than 1, and this implies that both the scattering angles $\theta_{\uparrow}^{A}$ and $\theta_{\downarrow}^{T}$ are well defined at all the injection angles $\theta_{\downarrow}$.
 
On the other side, when injecting spin-up particles, Andreev reflection and transmission in the superconducting layer can occur provided that the injection angle is less than the following critical values, respectively: %
\begin{eqnarray}
\theta^{c A}_{\downarrow} & = &\arcsin \left[\sqrt{\frac{1-X}{Y(1+X)}}\right]\\
\theta^{c T}_{\uparrow} & = & \arcsin \left[\sqrt{\frac{1}{\sqrt{Y}(1+X)}}\right] \; .
\label{CritAng}
\end{eqnarray}
In the considered two-dimensional limit, the magnetization of each ferromagnetic lead, defined as $M=(n_{\uparrow}-n_{\downarrow)}/(n_{\uparrow}+n_{\downarrow})$, with $n_{\uparrow (\downarrow)}$ being the number of spin-up (spin-down) electrons in the ferromagnetic layer, can be expressed as a function of the microscopic parameters $X$ and $Y$ defined in the main text as~\cite{Annunziata09}:
\begin{equation}
M =\frac{(X+1)Y -(1-X)}{(1+Y)+X(Y-1)} \; .
\end{equation}
Using this expression, it is possible to write down the limiting angle for the Andreev processes as
\begin{equation}
\theta^{c A}_{\downarrow} = \arcsin \left[\sqrt{\frac{1-M}{ 1+M}}\right] \; .
\end{equation}
\noindent This implies that the limiting angle for the Andreev reflections of spin-up electrons (and holes) does not depend on the mechanism responsible for the ferromagnetic order, but only depends on the value assumed by the magnetization, as we plot in Fig.~\ref{figthetaCRA}. On the other hand, the critical angle for the transmission in S crucially depends on the value of $X$ and $Y$, as shown in Fig.~\ref{figthetaCRT}. For any fixed value of the magnetization, it takes a smaller value in the case of the pure SMM mechanism compared to the Stoner one. However, the comparison between the critical angle for transmission into S  with that for the Andreev reflection shows that the actual limit to the injection cone of particles from one ferromagnetic lead to the other comes from the Andreev reflections, since the corresponding limiting angle is systematically smaller than that for the transmission into the S side, both for the Stoner [Fig.~\ref{figthetaCRAvsCRT}(a)] and for the SMM mechanism [Fig.~\ref{figthetaCRAvsCRT} (b)]. In principle, virtual Andreev reflections characterized by imaginary momenta could also be allowed for injection angles above $\theta^{c A}_\downarrow$, but in this case we have found no solution to our system of equations.

\section*{Appendix E}

Here we show that, differently from the Stoner case, for SMM ferromagnetic layers the probabilities corresponding to the different scattering processes at transparent interfaces ($\textsf{Z}=0$) and perpendicular injection direction ($\theta_{\sigma}=0$) are independent of the spin orientation of the injected particles. This feature comes from a symmetry between spin-up and spin-down carriers exhibited by the system of coupled linear equations (\ref{BC_system}), which only holds in the SMM case at $\textsf{Z}=0$ and $\theta_{\sigma}=0$. 
In the following we demonstrate it in the case where carriers are electrons, but it also holds in the hole case. To this purpose, we note that system (\ref{BC_system}) can be written in a compact form as
\begin{equation}
\hat{\textsf{M}}_{e \sigma} \textsf{X}_{e \sigma} = \textsf{Y}_{e \sigma} \; ,
\label{eqSYS}
\end{equation}
where $\textsf{X}_{e \sigma}$ is the vector of the unknown variables $\textsf{X}_{e \sigma}=(\textsf{a}_{e \bar{\sigma}}, \textsf{b}_{e \sigma}, \textsf{c}_{e \sigma}, \textsf{d}_{e \bar{\sigma}}, \alpha_{e \sigma}, \beta_{e \bar{\sigma}},\gamma_{e \sigma},\eta_{e \bar{\sigma}} ).$ 
In the case of injected electrons with spin $\sigma$, the matrix $\hat{\textsf{M}}_{e \sigma}$ and the vector $\textsf{Y}_{e \sigma}$ are
\begin{widetext}
\begin{align*}
\hat{\textsf{M}}_{e \sigma}= \begin{pmatrix}
  0 & -1 & 0 & 0 & u_0 & v_0 & u_0 & v_0 \\
 -1 &  0 & 0 & 0 & v_0 & u_0 & v_0 & u_0 \\
  0 &  0 & -e^{i \tilde{q}_{\sigma} l} & 0 & u_0 \ \Sigma_{e}^{+} & v_0\  \Sigma_{h}^{-} & u_0 \  \Sigma_{e}^{-} & v_0 \ \Sigma_{h}^{+} \\
  0 & 0 & 0 & -e^{-i \tilde{q}_{\bar\sigma} l} &  v_0 \ \Sigma_{e}^{+} & u_0 \  \Sigma_{h}^{-} & v_0 \ \Sigma_{e}^{-} & u_0 \ \Sigma_{h}^{+} \\
    0 &  i \textsf{Z}+\tilde{q}_{\sigma}/\sqrt{Y} & 0 & 0 & u_0 & -v_0 & -u_0 & v_0 \\
  -\tilde{q}_{\bar\sigma}\sqrt{Y} + i \textsf{Z} & 0 & 0 & 0 & v_0& -u_0& -v_0& u_0 \\
  0 &  0 & e^{i \tilde{q}_{\sigma}l}(\tilde{q}_{\sigma}/\sqrt{Y}+ i \textsf{Z}) &  0 & -u_0 \ \Sigma_{e}^{+} & v_0 \ \Sigma_{h}^{-} & u_0 \ \Sigma_{e}^{-} & -v_0 \  \Sigma_{h}^{+}\\
  0 & 0 &  0 & -e^{-i\tilde{q}_{\bar\sigma}l}(\tilde{q}_{\bar\sigma}\sqrt{Y}-i \textsf{Z})& -v_0 \ \Sigma_{e}^{+} & u_0  \ \Sigma_{h}^{-} & v_0 \ \Sigma_{e}^{-} & -u_0  \ \Sigma_{h}^{+} \\
 \end{pmatrix}
\end{align*}
\end{widetext}
and 
\[
\textsf{Y}_{e \sigma}=\left(1, 0, 0, 0, \tilde{q}_{\sigma}/\sqrt{Y} -i \textsf{Z}, 0, 0, 0\right) \; ,
\]
having defined $\Sigma_{e,h}^{\pm}=e^{\pm i \tilde{k}^{e,h}L}$ and $l=L \ k_F$.

For $\textsf{Z}=0$ and $\theta_{\sigma}=0$, the above matrices for the two spin species are related to each other through the following relations:
\begin{eqnarray}
\textsf{S} \hat{\textsf{M}}_{e\uparrow} & = &  \hat{\textsf{M}}_{e\downarrow} \, \textsf{U} \label{eqMdw1} \\
\textsf{S} \textsf{Y}_{e\uparrow} & = & \textsf{Y}_{e\downarrow} \label{eqMdw2} \; .
\end{eqnarray} 
Here $\textsf{S}$ is a off-diagonal block matrix $\textsf{S}= \eta_x \gamma_0$, $\gamma_0$ being one of the 4x4 Dirac matrices, and $\eta_x$ is a 8x8 matrix defined as 
\begin{eqnarray}
\eta_x=\left(
\begin{array}{cccc}
  0& \textsf{I}_4\\
  \textsf{I}_4 & 0
 \end{array}
\right) \; ,
\end{eqnarray}
$\textsf{I}_4$ is the 4x4 unit matrix and $\textsf{U}$ is a diagonal matrix defined as
\begin{equation}
\textsf{U}={\rm diag}\left(s^2,-1, e^{\imath l \Gamma}, -s^2 \ e^{\imath l \Gamma},s,-s,-s,s\right),
\end{equation}
with $\Gamma=\frac{-1+s^2}{s}$, and $ s=Y_1^{1/4}$. From Eqs.~(\ref{eqMdw1})-(\ref{eqMdw2}) it follows  that the unknown vector for injected spin-down electrons is linked to that for the injected spin-up ones through the relation: $\textsf{X}_{e\downarrow}=\textsf{U}\textsf{X}_{e\uparrow}$.  
Moreover, in the SMM case the probability vector defined as $\textsf{P}_{\sigma}=\left(\textsf{A}_{e\sigma},\textsf{B}_{e\sigma},\textsf{C}_{e\sigma},\textsf{D}_{e\sigma} \right)$, can be expressed in terms of the unknown coefficients vector $\tilde{\textsf{X}}_{\sigma}=\left(\textsf{a}_{e\bar{\sigma}},\textsf{b}_{e\sigma},\textsf{c}_{e\sigma}, \textsf{d}_{e\bar{\sigma}} \right)$ as reported in Eqs.~(\ref{eqApr})-(\ref{eqDpr}). In matrix form, we have
\begin{eqnarray}
\textsf{P}_{\sigma}=\tilde{\textsf{X}}_{\sigma}^{\dagger}\ \textsf{R}_{\sigma} \tilde{\textsf{X}}_{\sigma}
\end{eqnarray}
with 
\begin{eqnarray} 
\textsf{R}_{\uparrow}=
\left(
\begin{array}{ccccc}
s^2&0&0&0\\
0&1&0&0\\
0&0&1&0\\
0&0&0&s^2
\end{array}
\right) 
\end{eqnarray}
and $\textsf{R}_{\downarrow}=\left(\textsf{R}_{\uparrow}\right)^{-1}$. 
 
Using Eqs.~(\ref{eqSYS})-(\ref{eqMdw2}), we can write
\begin{eqnarray}
\textsf{P}_{\downarrow}&=& \tilde{\textsf{X}}_{\downarrow}^{\dagger}\ \textsf{R}_{\downarrow} \ \tilde{\textsf{X}}_{\downarrow}\\
&=& \left(\textsf{U} \ \tilde{\textsf{X}}_{\uparrow} \right)^{\dagger}   \textsf{R}_{\downarrow} \left(\textsf{U} \ \tilde{\textsf{X}}_{\uparrow}  \right)\\
&=&   \tilde{\textsf{X}}_{\uparrow} ^{\dagger} \left(\tilde{\textsf{U}}^{\dagger} \  \textsf{R}_{\downarrow} \tilde{\textsf{U}} \right)\ \tilde{\textsf{X}}_{\uparrow}   \\
&=& \tilde{\textsf{X}}_{\uparrow} ^{\dagger} \   \textsf{R}_{\uparrow}   \tilde{\textsf{X}}_{\uparrow} =\textsf{P}_{\uparrow}
\end{eqnarray}
with $ \tilde{\textsf{U}}^{\dagger}  \tilde{\textsf{U}}=  \left( \textsf{R}_{\uparrow}\right)^2$. This demonstrates the equality of the probability amplitudes for the injection of spin-up and spin-down electrons in the SMM case. The same holds also in the case of injected holes with opposite spin states.

\end{document}